\documentclass[journal,draftcls,onecolumn,12pt,twoside]{IEEEtranTCOM}

\normalsize

\usepackage{graphicx}
\usepackage{cite}
\usepackage{subfigure}
\usepackage{multirow}
\usepackage{url}
\usepackage{setspace}

\ifCLASSINFOpdf
\else
\fi

\usepackage{amsfonts,amssymb}
\usepackage{amsmath}
\usepackage{comment}
\usepackage{pgfplots}

\renewcommand{\baselinestretch}{1.0}

\IEEEoverridecommandlockouts

\newcommand{\bpara}[1]{\noindent\textbf{#1}~}

\begin{document}
%
\title{PABO: Mitigating Congestion via Packet Bounce in Data Center Networks\thanks{A preliminary version of the paper has appeared at IEEE ICC 2017 \cite{Shi-PABO-2017}. The Corresponding Author is Zhiyong Liu (zyliu@ict.ac.cn). This work was partially supported by the National Key Research and Development Program of China (grant number 2017YFB1010001), the National Natural Science Foundation of China (grant numbers 61520106005, 61761136014), and the Collaborative Research Center 1053 (MAKI) of the German Research Foundation (DFG). 
}}
%
%
%

\author{Xiang~Shi, 
        Lin~Wang,~\IEEEmembership{Member,~IEEE}, 
        Fa~Zhang,
        Kai~Zheng,~\IEEEmembership{Senior Member,~IEEE}, 
        Max~M\"uhlh\"auser,
        and~Zhiyong~Liu
        
\thanks{X. Shi, F. Zhang, and Zhiyong Liu are with the High Performance Computer Research Center, Institute of Computing Technology, Chinese Academy of Sciences, Beijing 100101, China. X. Shi is also with University of Chinese Academy of Sciences, Beijing, China. This work was done while X. Shi was visiting Technische Universit\"at Darmstadt.}
\thanks{L. Wang and M. M\"uhlh\"auser are with the Telecooperation Lab at Technische Universit\"at Darmstadt, Germany.}
\thanks{K. Zheng is with Huawei Technologies, Beijing, China.}
}

%
%

\markboth{COMPUTER COMMUNICATIONS}%
{Submitted paper}
%



\maketitle

\begin{abstract}

In today's data center, a diverse mix of throughput-sensitive long flows and delay-sensitive short flows are commonly presented.
However, commodity switches used in a typical data center network are usually shallow-buffered for the sake of reducing queueing delay and deployment cost.
The direct outcome is that the queue occupation by long flows could potentially block the transmission of delay-sensitive short flows, leading to degraded performance. Congestion can also be caused by the synchronization of multiple TCP connections for short flows, as typically seen in the partition/aggregate traffic pattern. The congestion is usually transient and any end-device intervention through the timeout-based pathway would result in suboptimal performance.
While multiple end-to-end transport-layer solutions have been proposed, none of them has tackled the real challenge: reliable transmission in the network. In this paper, we fill this gap by presenting PABO -- a novel link-layer design that can mitigate congestion by temporarily bouncing packets to upstream switches. PABO's design fulfills the following goals: $i)$ providing per-flow based flow control on the link layer, $ii)$ handling transient congestion without the intervention of end devices, and $iii)$ gradually back propagating the congestion signal to the source when the network is not capable to handle the congestion. We present the detailed design of PABO and complete a proof-of-concept implementation. We discuss the impact of system parameters on packet out-of-order delivery and conduct extensive experiments to prove the effectiveness of PABO. We examine the basic properties of PABO using a tree-based topology, and further evaluate the overall performance of PABO using a realistic Fattree topology for data center networks. Experiment results show that PABO can provide prominent advantage of mitigating transient congestions and can achieve significant gain on flow completion time.
\end{abstract}


%
\IEEEpeerreviewmaketitle

\section{Introduction}

Nowadays, organizations like large companies or universities built various sizes of data centers for different purpose today. To interconnect all of the data center resources together, a dedicated network (a.k.a Data Center Network, DCN) is used to support high-speed communications between servers with high availability in data centers.
For reasons of queueing delay and deployment cost, DCN are usually composed of shallow-buffered commodity switches at low costs.
However, data center applications such as web search, recommendation systems and online social networks can generate a diverse mix of short and long flows, demanding high utilization for long flows, low latency for short flows, and high burst tolerance \cite{Alizadeh-DCTCP-2010}.
Long flows lead to queue buildup in switches, which reduces the amount of buffer space available for delay-sensitive short flows and burst traffic, leading to frequent packet losses and retransmissions.
Meanwhile, data center applications such as real-time applications and data intensive applications produce traffic that follows the partition/aggregate pattern, overflowing the bottleneck switch buffer in a short period of time.
The final performance of the partition/aggregate pattern is determined by the slowest TCP connection that suffers timeout due to packet losses.
Therefore, for short flows and bursty traffic that are delay-sensitive, even a few lost packets can trigger window reduction and retransmission timeout, causing crucial performance degradation and high application latencies.

It has been demonstrated by \cite{Alizadeh-DCTCP-2010} that the greedy fashion of the traditional TCP and its variants fail to satisfy the performance requirements of these flows and thus, various TCP-like protocols such as DCTCP \cite{Alizadeh-DCTCP-2010} and ICTCP \cite{Wu-ICTCP-2010} have been proposed dedicatedly for DCN environments. However, none of the proposals can guarantee one hundred percent prevention of packet losses or timeouts \cite{Zarifis-DIBS-2014} -- the main culprit for performance degradation in DCNs, leaving poor performance in most cases.

%
%
%
%
The need of end-to-end solutions (on the transport layer) for reliable data transmission comes from the fact that reliable point-to-point transmission mechanisms (on the data link layer) are not available in the current protocol stack. 
If packet processing in a link is slower than packet arrival, excessive packets competing for the same output port of the switch will lead to queue buildup. Continuous queue buildup will overflow the output buffer, then the subsequently arrived packets will be dropped. Neither the source nor the destination will be explicitly notified about this congestion and thus will have no knowledge of when and where packet losses have happened. The dropped packets will be retransmitted by upper-layer congestion control protocols (e.g., TCP, DCCP), typically through a timeout-based pathway.

As one of the very few proposals toward the goal of providing reliability on the link layer, PAUSE Frame \cite{Feuser_PFC_1999} allows a switch to send a special frame (namely PAUSE frame) to its upstream switches, which results in a temporary halt of the data transmission. However, all the flows on the same link will be affected without considering their contributions on the congestion. To alleviate this situation, PFC (Priority Flow Control) \cite{Feuser_PFC_1999} further extends the idea to support eight service classes and consequently, PAUSE frames can be sent independently for each service class. Despite the lack of per-flow control, the parameters in PFC have to be carefully tuned individually according to each network circumstance in order to guarantee congestion-free \cite{Zarifis-DIBS-2014}. 

In this paper, we propose PAcket BOunce (PABO), a novel link-layer protocol design that can provide reliable data transmission in the network. Instead of dropping the excess packets when facing buffer saturation, PABO chooses to bounce them back to upstream switches. On the one hand, transient congestion can be mitigated at a per-flow granularity, which can help achieve significant performance gain for short-lived flows and burst flows as in the typical incast scenario \cite{Chen-Incast-2009} in a DCN. On the other hand, the congestion is gradually back propagated toward the source and can finally be handled by the source if the congestion cannot be solved right in the network. To the best of our knowledge, PABO is among the first solutions for reliable transmission and in-network congestion mitigation.


Our contributions can be summarized in five aspects: $\mathit{i}$) We propose a novel link-layer protocol, PABO, that supports full reliability and can handle transient congestions in the network. $\mathit{ii}$) We present the design of PABO and explain its components in detail. $\mathit{iii}$) We complete a proof-of-concept implementation of PABO in OMNeT++ \cite{OMNeT}.
$\mathit{iv}$) We investigate into the impact of PABO on the level of packet out-of-order, based on which we provide some insights for configuring PABO. $\mathit{v}$) We carry out extensive experiments to validate the basic properties of PABO using a tree topology and to evaluate PABO's overall performance using a realistic Fattree topology. 

The rest of the paper is organized as follows: Section II describes the design rationale of PABO and details its components one by one. Section III discusses the implementation of PABO. Section IV analyzes the relationship between PABO and the level of packet out-of-order. Section V and VI present the experimental results. Section VII summarizes related work and Section VIII concludes the paper.

\section{PABO's Design}

PABO is a link-layer solution for congestion mitigation based on back-propagation. 
Particularly, it is well suited for data center environments, where transient congestions (e.g., incast congestion) are commonly presented \cite{Kandula_TrafficPattern_2009,Kandula_Flyways_2009}. In addition, upper-layer congestion control protocols can be incorporated to achieve smooth congestion handling in all circumstances by taking advantage of the back propagation nature of PABO. 

\subsection{An Example}


\begin{figure}
\centering
\includegraphics[scale=0.4]{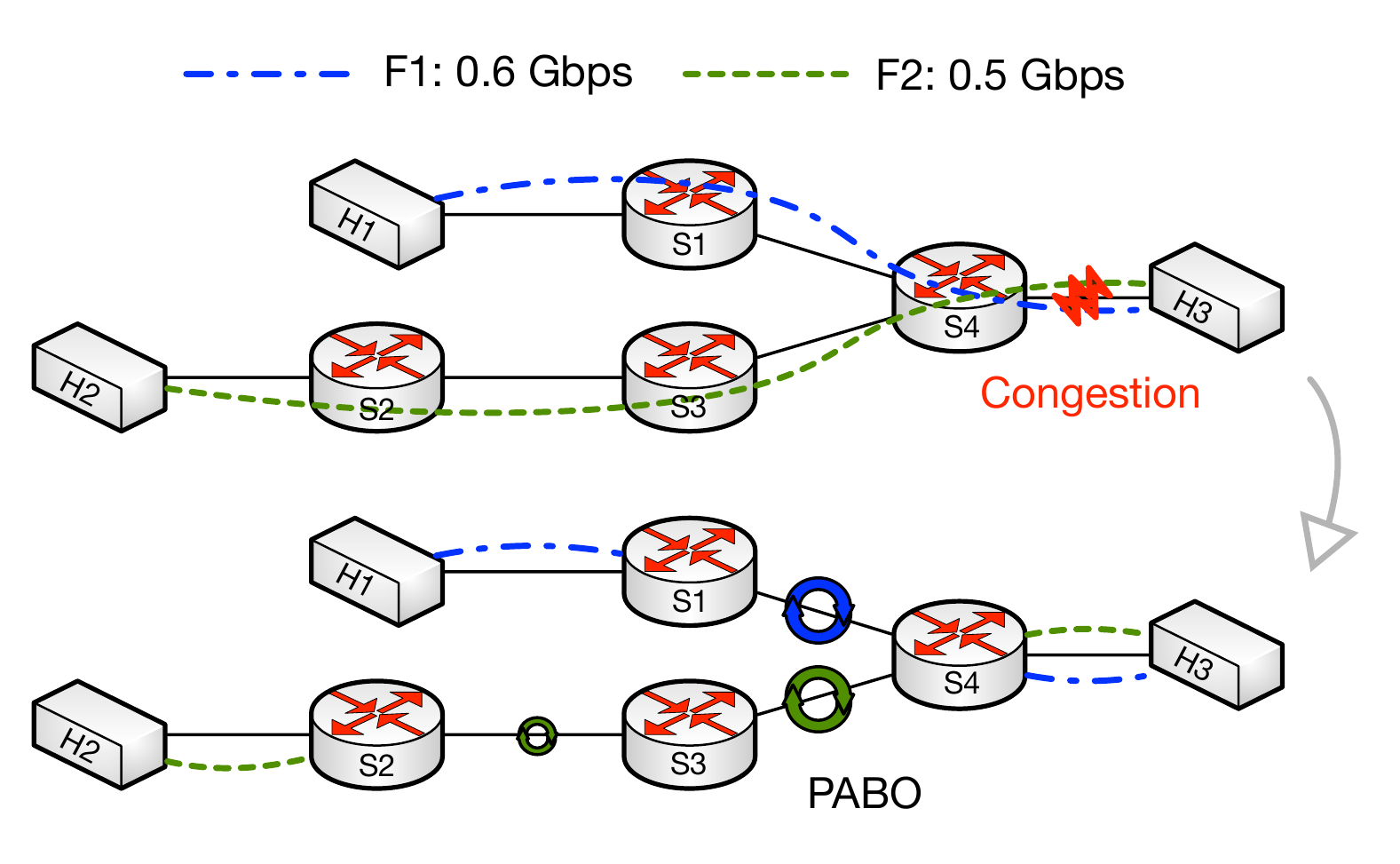}
\caption{\label{fig:example} A motivating example to demonstrate the idea of PABO, i.e., solving network congestion without dropping packets. We assume GigE links in the network.}
\vspace{-0.5cm}
\end{figure}


We provide a motivating example and explain why PABO is superior in handling congestion by providing reliable transmission in communication networks. Assume a GigE network and consider the scenario in Figure~\ref{fig:example}, where two short flows F1 (with rate $0.6$ Gbps) and F2 (with rate $0.5$ Gbps) that consist only tens of packets are destined for the same host H3. When congestion occurs on the link S4 - H3, PABO will bounce some of the packets to upstream switches (i.e., S1 and S3, respectively). The number of bounced packets will depend on the congestion level. Packet bouncing can be contagious reversely along the forwarding path. e.g., from S3 to S2. The bounced packets will then be forwarded to H3 again. Consequently, there will be packets bouncing back and forth on the few links in the network until the congestion vanishes.



\subsection{Overview}

\begin{figure}
\centering
\includegraphics[scale=0.4]{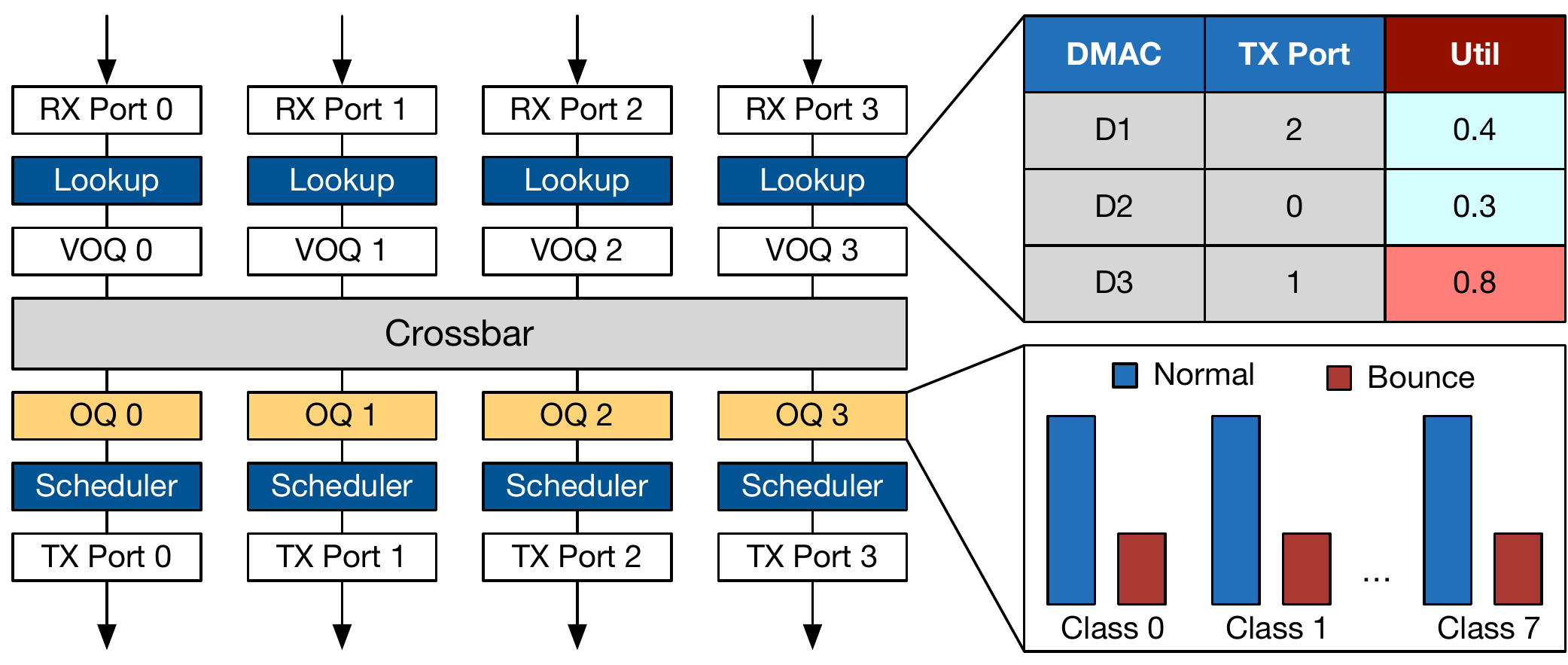}
\caption{\label{fig:switch-arch}An overview of PABO's design based on a Combined Input and Output Queue (CIOQ) switch model.}
\vspace{-0.5cm}
\end{figure}

We assume a general Combined Input and Output Queue (CIOQ) switch model \cite{Kesselman-CIOQ-2006} as depicted in Figure~\ref{fig:switch-arch}. This simplified switch model contains following modules: the FIB (Forwarding Information Base, also known as forwarding table) which contains MAC address-to-port mappings obtained from MAC address learning, the lookup unit, the virtual output queues, the output queues, and the output schedulers. 
Upon the arrival of a packet from any input interface, the packet will first go through the lookup unit. The lookup unit decides the output port for the packet by querying the FIB. Next, the packet will be passed into the corresponding virtual output queue (illustrated as VOQ in Figure~\ref{fig:switch-arch}). The packet will then be sent to the corresponding output queue through the crossbar and finally to the output interface by the output scheduler. Our design of PABO involves modifications on the following components of traditional switches: the lookup unit, the FIB, the output queue and output scheduler, and packet structure. We will discuss them one by one in the following subsections.

\subsection{Lookup Unit}
By querying the FIB, the lookup unit obtains the output port number for a packet corresponding to the packet's destination. Instead of forwarding the packet directly to its destined port, PABO introduces a probabilistic decision making process to decide where to forward the packet. Upon packet arrival, the lookup unit calculates a probability with which the packet will be bounced to its previous-hop switch. The calculation is based on a probability function $P$, which follows the principles:\\
\textbf{PRCP-1:} The probability should be zero when the queue is almost free and should be one when the queue is full;\\
\textbf{PRCP-2:} The probability should increase super-linearly with the queue utilization at the early stage of bouncing back to prevent the queue from fast buildup;\\
\textbf{PRCP-3:} Packets that have already been returned should receive smaller probability to be bounced back again.


To satisfy PRCP-3, we first define a base probability factor $P_b$ for each packet according to the number of times the packet has been bounced, i.e., $n_p$. When $n_p$ grows, the base probability factor should decrease dramatically to intentionally reduce the chance of a packet being persistently bounced, as this could result in a large delay. To this end, we use an exponential decay function in the following form for the base probability factor:

\begin{equation}
\label{eq:base_prob}
P_b(n_p) = { e^{\frac{\lambda}{n_p+1}}}
\end{equation}
where $\lambda>0$ is the exponential decay constant. It is always true that $P_b(n_p) > 1$ for any $n_p \in \mathbb{Z}_0^+$.

\begin{figure}[!t]
\centering
\subfigure[]{
\label{fig:prob_func1}
\includegraphics[scale=0.3]{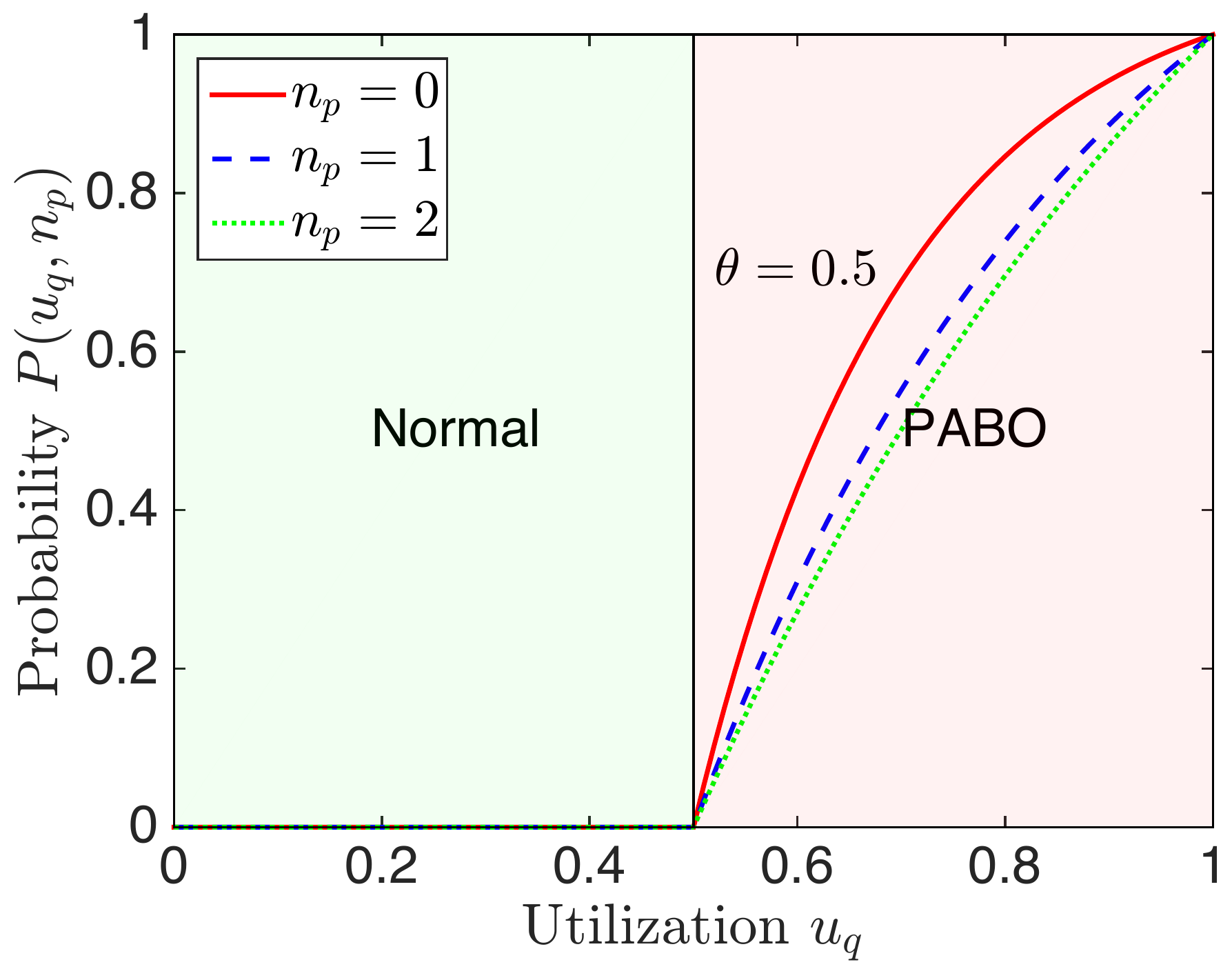}
}
\hspace{-0.15cm}
\subfigure[]{
\label{fig:prob_func2}
\includegraphics[scale=0.3]{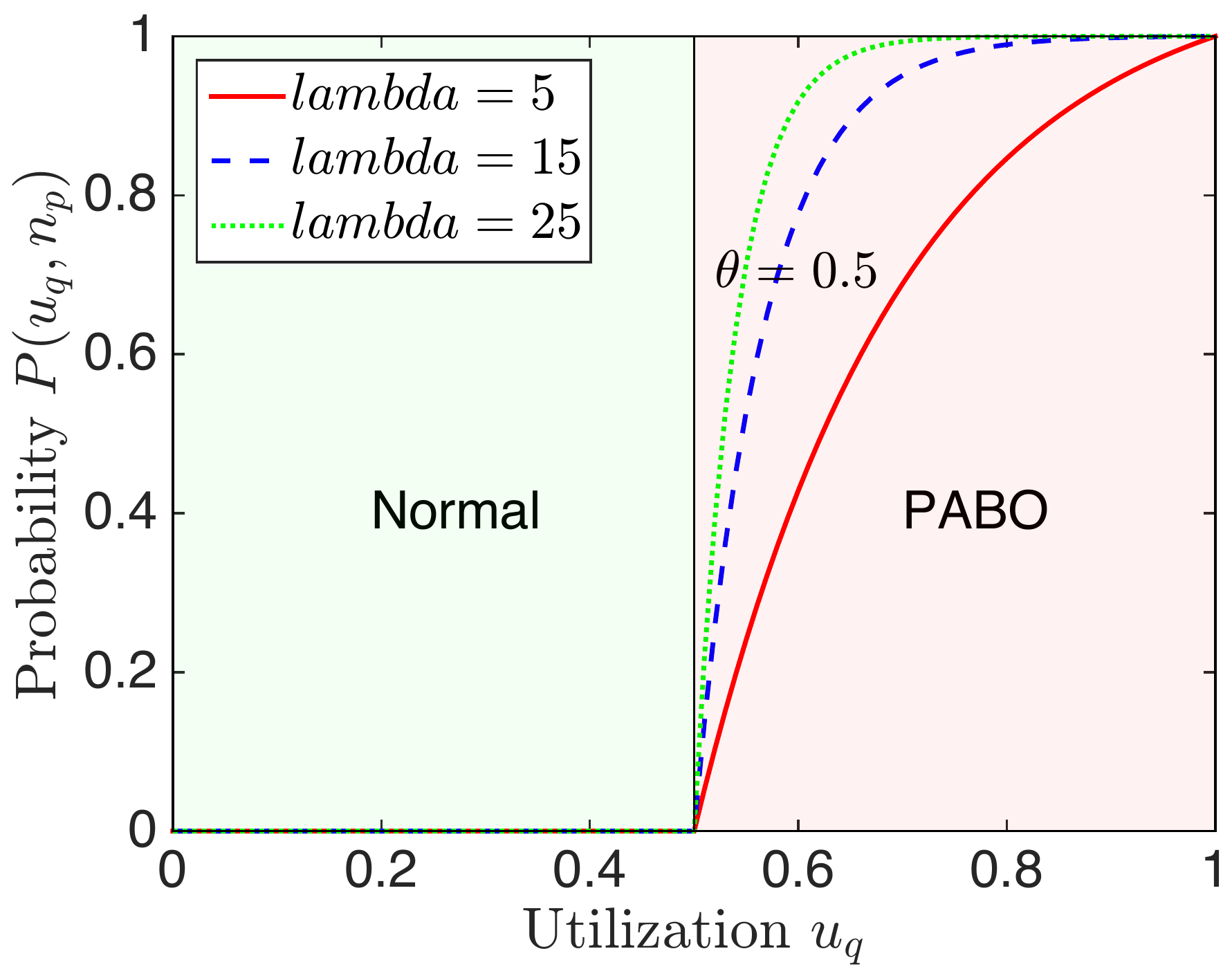}
}
\vspace{-0.3cm}
\caption{Probability function $P(u_q,n_p)$ for packet bounce decision making. We set the threshold $\theta$ to $0.5$: (a) we set the constant $\lambda$ to $5$, and we show the curves in cases of $n_p= 0, 1, 2$, respectively, and (b) we show the curves with $lambda= 5, 15, 25$ in the case of $n_p= 0$.
When $u_q \leq 0.5$, switches forward packets normally; PABO is only involved when $u_q > 0.5$.}
\label{fig:prob_func} 
\vspace{-0.7cm}
\end{figure}

We introduce a lower threshold $\theta \in [0, 1)$ for the output queue utilization $u_q$ and define $P(u_q, \cdot) = 0$ if $0 \leq u_q \leq \theta$ and $P(u_q, \cdot) = 1$ if $u_q = 1$. The first equation means that if the output queue utilization is under the predefined threshold, i.e., the queue is underutilized, there is no need to bounce packets; the second equation guarantees that when the output queue is full or overflows, all the upcoming packets have to be bounced back in order to avoid packet drops. These two equations ensure the validity of PRCP-1. Then we define for $\theta < u_q < 1$,
\begin{equation}
P(u_q, \cdot) = \alpha \cdot P_b^{-u_q} + \beta
\end{equation}
where $\alpha$ and $\beta$ are constants. Noting that $P(u_q, \cdot)$ also satisfies $(\theta, 0)$ and $(1, 1)$, we have
\begin{align}
\alpha = \frac{P_b^{\theta}}{P_b^{\theta-1} - 1}, 
\beta = \frac{1}{1-P_b^{\theta-1}}.
\end{align}
Substituting $P_b$ with Equation~(\ref{eq:base_prob}) 
and combining all the above cases, we have the closed form of $P$ satisfying PRCP-2 as
\begin{equation}
\label{eq:prob_func}
P(u_q, n_p) =\left\{
\begin{aligned}
&0 && 0 \leq u_q \leq \theta, \\
&\frac{{e^{\frac{\lambda(\theta-u_q)}{n_p+1}}}-1}{ e^{\frac{\lambda(\theta-1)}{n_p+1}}-1} && \theta < u_q \leq 1. \\
\end{aligned}
\right.
\end{equation}
The curves of function $P$, with $\theta = 0.5$, under different utilization $u_q$ are illustrated in Figure~\ref{fig:prob_func}. Note that $\theta$ has a major impact on the proportion of packets that will be bounced back at a given queue, while $\lambda$ can be used to roughly control the maximum number of hops each bounced packet will traverse. We will further verify these correlations in Section VI.

\subsection{FIB}

The FIB maintains mappings between the packet destination MAC address and the corresponding output port (i.e., interface) that the packet should be forwarded to. In PABO's design, the lookup unit relies on two parameters, $u_q$ and $n_p$, to make the forwarding decision for each packet as we just discussed. While $n_p$ can be obtained from the packet as we will describe in Section~\ref{subsec:packet}, $u_q$ needs to be available after the inquiry to the FIB from the lookup unit. To achieve this, we introduce an extra column named ``Util" in the forwarding table, as shown in Figure~\ref{fig:switch-arch}. In addition to maintaining the MAC-port mappings, the FIB also monitors and updates the output queue utilization $u_q$ for each output port. When a packet is forwarded to an output queue, the utilization $u_q$ of the queue will be updated by the following equation $u_q \leftarrow u_q + 1/C_q$, where $C_q$ is the maximum capacity of the queue. When a packet is expelled by the scheduler at an output queue, the corresponding value of $u_q$ in the FIB is also updated according to the following equation $u_q \leftarrow u_q - 1/C_q$.


\subsection{Output Queue}

We separate the bounced packets from the normal packets by assigning the bounced packets higher priority at the output queue (illustrated as OQ in Figure~\ref{fig:switch-arch}). This is due to the observation that compared to normal packets: packets that have already been bounced should be processed earlier as they have already been delayed during the bouncing process. To this end, we introduce two virtual sub-queues for the output queue, namely bounce queue and normal queue (for every service class). The packets in the bounce queue will enjoy higher priorities when being scheduled by the output scheduler. For simplicity, we adopt a straightforward scheduling strategy and we modify the output scheduler such that packets from the normal queue will be transmitted only if the bounce queue is empty. In a sequel, the bouncing delay will be compensated by reducing their queueing delay during their retry process.

\subsection{Packet}
\label{subsec:packet}

Each packet in the network will carry a counter $n_p$ to indicate how many times the packet has been bounced back. 
Take again the example in Figure~\ref{fig:example}. When congestion occurs at $S_4$-$H_3$, suppose the last normally reached switch for the packets from both flows $F_1$ and $F_2$ is $S_4$. For packets from flow $F_2$ that are bounced back to $S_3$ by $S_4$, the value of $n_p$ will be increased by one. When the bounced packets are forwarded out normally to $S_4$, the value of $n_p$ will stay the same. If these packets are bounced from $S_4$ to $S_3$ again, $n_p$ will increase to record the new bounce behaviors. 
This counter $n_p$ will be used by the lookup unit in switches as an input for the bounce probability calculation. By carefully setting the probability functions as we have already discussed, the probability that a packet is bounced back consistently for multiple times will be significantly reduced.


\subsection{End-host Support}

When persistent congestion occurs, there will be packets bounced straight back to the source along the reverse direction of the forwarding path. While end-host involvement is required to support this circumstance, the bounced packets can also serve as a congestion notification for upper-layer congestion control protocols. When receiving a bounced packet, the source will be notified that the congestion has happened along the forwarding path, and exceeds the ability of the network to handle it. In such cases, the source would reduce its sending rate. The extent of this rate adjustment will depend on the severity of the congestion, measured by for example the number of bounced packets the source has received during a certain amount of time. We claim that more sophisticated transport layer congestion control protocols can also be incorporated to further handle the congestion smoothly. The bounced packets received by the source will be injected into the source's output queue again for further retransmission, which also ensures that no packets will be dropped.




\section{Implementation}


To validate the effectiveness of PABO, we completed a proof-of-concept implementation based on the INET framework for OMNeT++ \cite{OMNeT}. Our implementation code is open-sourced at \cite{PABO-GH}. By overriding the corresponding link-layer modules, we created Ethernet switches and host models that can support PABO. 
The detailed modifications made to each module will be explained in the following.

\bpara{EtherSwitch.}  
We mainly modify the implementation of its submodules including $\mathtt{MACTable}$, $\mathtt{MACRelayUnit}$, $\mathtt{EtherQosQueue}$. We introduce a float-point variable for each entry in the $\mathtt{MACTable}$ module to keep track of the output queue utilization at the corresponding output interface. In $\mathtt{MACRelayUnit}$, We alter the implementation of the forwarding strategy (i.e., function $\mathtt{handleAndDispatchFrame}$) by applying our probability-based forwarding decision making mechanism. The utilization variable in the $\mathtt{MACTable}$ is also updated accordingly after a forwarding decision has been made. Furthermore, we disable the MAC address self-learning process when receiving bounced packets, as the destination addresses of bounced packets are already in the forwarding table. $\mathtt{EtherQosQueue}$ is a typical buffer module type, which is composed of classifier, queue, and scheduler. In addition to the default $\mathtt{dataQueue}$ (as $\mathtt{normalQueue}$), we introduce another queue called $\mathtt{bounceQueue}$. We alter the classifier in order to separate bounced packets from normal ones. Bounced packets are stored in $\mathtt{bounceQueue}$, while normal packets are sent into $\mathtt{normalQueue}$. The $\mathtt{normalQueue}$ and $\mathtt{bounceQueue}$ are with the $\mathtt{DropTailQueue}$ type. Finally, we modify the output scheduler $\mathtt{PriorityScheduler}$ where we give priorities to the packets in the $\mathtt{bounceQueue}$ to reduce delay. 

\bpara{EtherFrame.} This is a message type representing link-layer frame. It contains the common header fields and payloads. To keep track of the value of $n_p$, i.e., the number of times the packet has been bounced back, we add a non-negative integer counter $\mathtt{bouncedHop}$ in the header. This counter will increase by one every time the packet is bounced back by one hop and will stay the same if the packet is forwarded normally. 
Then we introduce three other parameters for further analysis. We introduce another counter $\mathtt{bouncedDistance}$ for each packet to indicate how far (measured by the number of hops) it has been bounced from its last normally reached switch before it was first bounced. This counter will increase by one if the packet is bounced and will decrease by one if the packet is normally forwarded.
For each packet, to record the farthest distance it has been bounced, we introduce the parameter $\mathtt{maxBouncedDistance}$.
Meanwhile, we also add a non-negative integer counter $\mathtt{totalHop}$ to record the total number of hops (normal plus bounced, including the sender) that the packet has traversed in the network.


\bpara{Host.} The host is $\mathtt{StandardHost}$ type -- an example host contains modules related to link layer, network layer, transport layer and application layer. We mainly make modifications to the link layer of the $\mathtt{StandardHost}$ to generate our host model.
 Ideally, PABO is so far only used for mitigating transient congestion in the network without the involvement of end-hosts. However, it is possible that the congestion condition persists too long and the bounced packets will finally reach the sender. To handle this situation, we modify the $\mathtt{EtherMAC}$ module in the $\mathtt{StandardHost}$ to avoid dropping bounced packets that are not destined for this host. Then we modify the $\mathtt{EtherEncap}$ module to check whether there is bounced packet or not. If so, the bounced packets will be sent to the sender's buffer for retransimission. Note that the type of sender's buffer is also $\mathtt{EtherQosQueue}$. As a result, the same modifications we made for switches can also be applied for $\mathtt{EtherHost}$. 

In addition to the PABO implementation, we make some special modifications to the transport layer of the $\mathtt{StandardHost}$ to measure the level of packet out-of-order.
In the TCP sender, we record the sending order of each packet into a special queue called $\mathtt{sentSeqQueue}$.
We maintain a send counter in the TCP sender. Every time a new packet is sent, the packet is labeled with a unique sending order $s_i$, and this mapping information (packet $i$, $s_i$) will be recorded into $\mathtt{sentSeqQueue}$ until the packet $i$ is ACKed. If packet $i$ is retransmitted, the sender will look up in $\mathtt{sentSeqQueue}$ to find the corresponding $s_i$, and label the retransmitted packet with it.
In the TCP receiver, we maintain a reception counter. Each received packet is assigned a receiving order $r_i$ (loss and duplicate packets are ignored). For each packet, we calculate the difference between its $s_i$ and $r_i$. Finally, we can get a distribution of displacement of packets, which will be used for further analysis.

\section{Packet Out-of-Order Analysis}
\label{sec:out-of-order}

Packets arriving at receiver disobeying their sending order is called packet out-of-order delivery \cite{Leung-PacketReordering-2007}. 
TCP requires the in-order of packets, which means the receiving order of the packets is supposed to be the same as their sending order. 
Out-of-order packets will generate sequence holes on the receiver side and then, TCP will duplicate the ACK to request the missing packets. Three continuous duplicate ACKs brought by out-of-order packets can trigger spurious fast retransmission \cite{Ludwig-Eifel-2000}, bringing unnecessary packet delay as well as reduced congestion window. 
Studies have shown that packet out-of-order delivery is not rare \cite{Bennett-Pathological-1999, Paxson-Dynamics-1999} and it can be caused by multiple factors such as packet loss, parallelism within routers or switches, different path delays in packet-level multi-path routing, route fluttering and so on \cite{Leung-PacketReordering-2007}.
To deal with this problem,  \cite{Zhang-ReorderingTCP-2003} increases the number of duplicate acknowledgments required to trigger fast retransmission and  \cite{Alizadeh-DisableFRTCP-2013} disables fast retransmissions entirely.

In our case, PABO allows packets to be retuned on the forwarding path in order to guarantee no packet loss. On the other hand, this bouncing behavior will disturb the packet forwarding direction, thus inevitably leading to increased number of out-of-order packets.
This effect on packet out-of-order is also a congestion indication that PABO tries to convey to TCP.
If the number of duplicate ACKs grows to a certain threshold, TCP can finally handle the congestion by suppressing the sending rate. 
However, window reduction should be avoided in transient congestions as it can result in unnecessary performance damage, while things can be solved right by using PABO instead.
In this section, we will dissect the impact of each of the system parameters (e.g., $\theta$ and $\lambda$ in the bounce probability function~(\ref{eq:prob_func})) to packet out-of-order. From the analysis,  we provide insights focusing on  PABO configuration.

\subsection{Measuring Packet Out-of-Order}
To measure the level of packet out-of-order delivery in a packet flow, various methods have been proposed. Among them, Reorder Density (RD) captures packet out-of-order delivery by a weighted distribution of the displacement of packets \cite{Jayasumana-RD-2008}. Consider a sequence of packets sent in the order of $[1,2,...,N]$, which is referred to as sending index $s_i$ for packet $i$. When arriving at the receiver side, each packet $i$ will be assigned a receiving index recording the order of reception, denoted by $r_i$. The difference between the sending index and the receiving index, denoted by $d_i$ for each packet $i$, is calculated as
\begin{equation}
\label{eq:displacement}
d_i = r_i - s_i.
\end{equation}
If $d_i>0$, packet $i$ is considered to be late; $d_i<0$ means that packet $i$ arrives earlier than expected; $d_i=0$ means there is no out-of-order event occurred. RD also introduces a threshold $D_T>0$ on $|d_i|$, beyond which an early or a late packet is deemed lost. Lost or duplicated packets will not be assigned any receive index. Then, we define distribution vector $S[k]$ which contains the number of packets with a displacement of $k$. By normalizing it to the total number of non-duplicated received packets $N$, we obtain the following weighted distribution of displacements, denoted by $RD[k]$.

\begin{equation}
\label{eq:weighted_distribution_displacement}
RD[k] = \frac{S[k]}{N},    -D_T \leq k \leq D_T.
\end{equation}
Based on RD, we now derive a new metric called \emph{reorder entropy} to quantitatively analyze the properties of the distribution \cite{Ye-ReorderEntropy-2006}.
Reorder entropy uses a single value to characterize the level of out-of-order in a packet flow, reflecting the fraction of packets displaced and the severity of packet displacement. The formal definition of reorder entropy is given by
\begin{equation}
\label{eq:reorder_entropy}
E_R = (-1) \cdot \sum_{i=-D_T}^{i=D_T}(RD[i]  \cdot  \ln{RD[i]}).
\end{equation}
It can be verified that larger reordering entropies represent a more dispersed distribution of packet displacement, translating into a more severe packet out-of-order event. If there is no packet out-of-order at all, the reorder entropy should be equal to zero.

\subsection{Impact of PABO on Reorder Entropy}

To explore the impact of PABO on packet out-of-order, we conduct some experiments to investigate how the reorder entropy changes with different PABO parameters. In particular, we tune the values for parameters $\theta$ and $\lambda$ in the bounce probability function as in Equation~(\ref{eq:prob_func}) and we report our major observations, based on which we discuss possible ways to improve PABO in terms of packet out-of-order.

\begin{figure}
\centering
\includegraphics[scale=0.6]{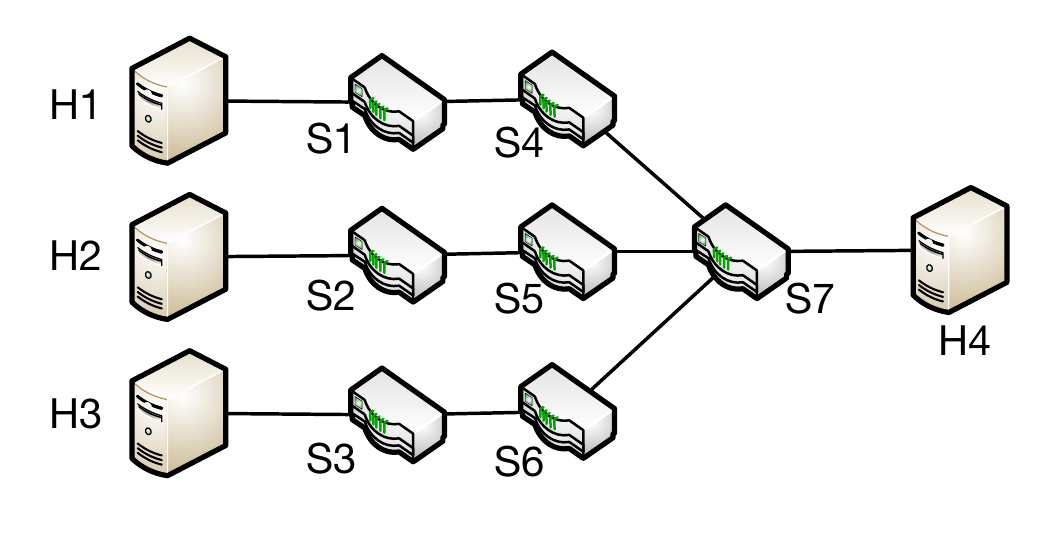}
\vspace{-0.3cm}
\caption{The network topology used for investigating the impact of PABO on the level of packet out-of-order, and also for conducting the hop-by-hop evaluation of PABO.}
\label{fig:exp-topo}
\vspace{-0.5cm}
\end{figure}

\bpara{Simulation Setup.}
We adopt a tree-based network topology consisting of three servers (i.e., H1, H2 and H3) and one client (i.e., H4) connected by seven switches, as depicted in Figure~\ref{fig:exp-topo}. All the links in the network are assumed to have the same rate of 1Gbps.
We consider a scenario where the client establishes and maintains three concurrent TCP connections with the three servers, respectively.
The topology shows that the client is three-hop away from each of the servers and the data flows from all the servers will be aggregated at the last-hop switch directly connected to the client. 
All the experimental results presented in this section will be combined measurement results of the three concurrent TCP connections.

\bpara{Traffic.}
We create communication patterns in the considered scenario to simulate the expected congestion conditions. 
In our experiment, we set up one TCP session for each of the TCP connections during the whole simulation time. Each of the TCP session includes four TCP requests, in which the hosts behave in a request-reply style: The client sends a request (200B) with the expected reply length (1MiB) to the server, then the server responds immediately with the requested length of data. 
Each TCP request represents an appearance of transient congestion at switch S7 due to the fact that all the three servers will send data to the client in a synchronized fashion. 
This setting simply emulates the partition/aggregate traffic pattern that is very popular in a data center network and can be conveniently monitored.
In the rest of the paper, we will refer to it as the partition/aggregate traffic for ease of expression.
The time gap between TCP requests is set to one second, which is enough to avoid overlaps between the periodic congestion appearances. To simulate a transient congestion (e.g., incast congestion), we set buffer sizes to be small and flow rates to be bursty. More specifically, the queue capacities of both $\mathtt{dataQueue}$ and $\mathtt{bouncedQueue}$ in all the switches and hosts are set to 100. To remove the limit on flow rate by the congestion control mechanism in TCP, we set $\mathtt{ssthresh}$ to be arbitrarily high so that the servers will perform slow start without the limit of $\mathtt{ssthresh}$. At the same time, the advertised window of the client is set to 45535 bytes to enable the growth of flow rate. We also disable fast retransmission in TCP and set the retransmission timeout (RTO) to its upper bound 240s \cite{Braden-RFC-1989}. This way, the packet out-of-order caused by packet retransmission is eliminated so we can observe the packet out-of-order brought only by the bouncing behavior of PABO.

\begin{figure}
\centering
\includegraphics[scale=0.4]{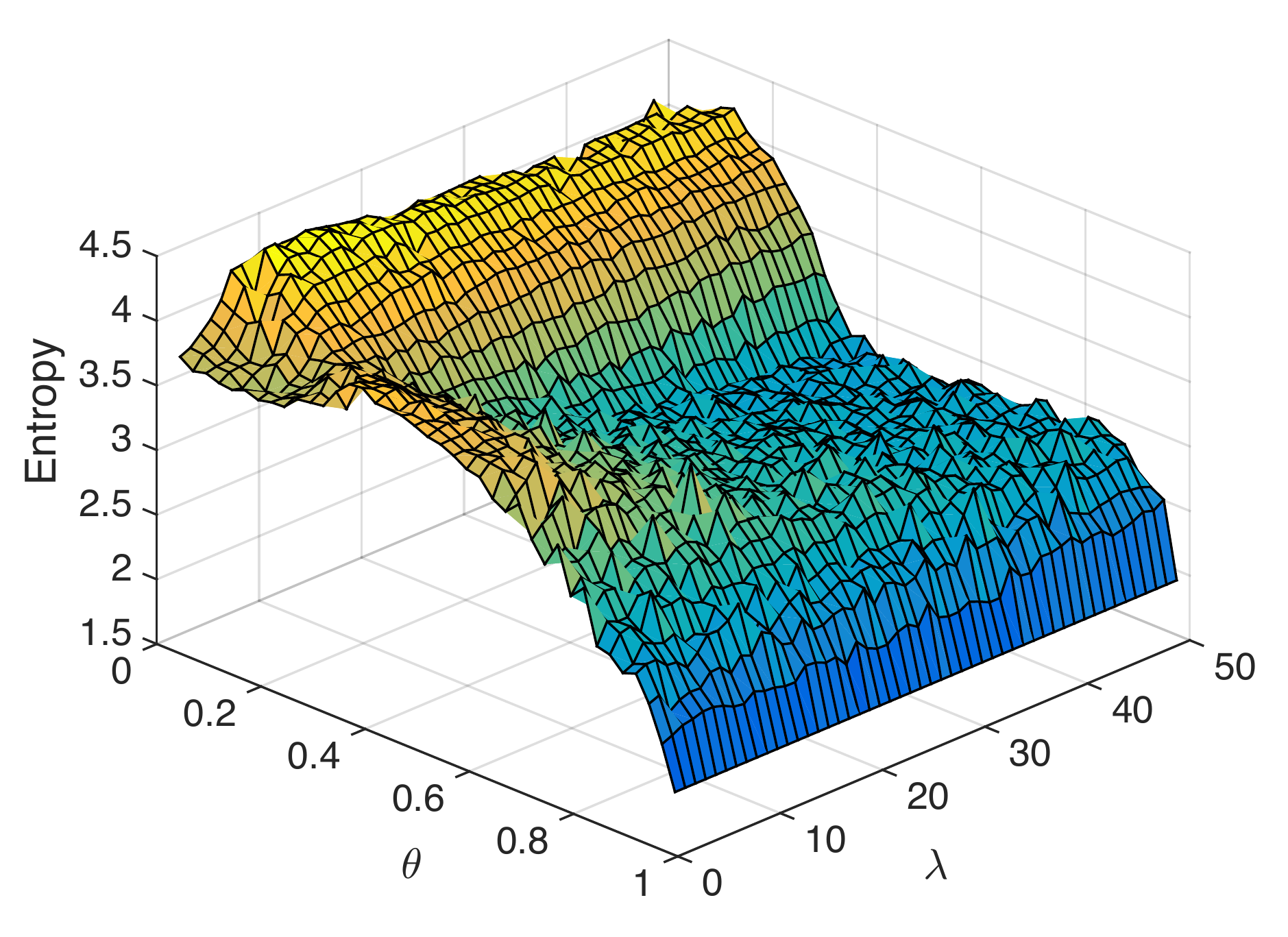}
\vspace{-0.3cm}
\caption{The relationship between $\theta$, $\lambda$ in bounce probability function and entropy. Darker color of a grid means smaller entropy. In general, the change of $\theta$ has  greater influence on entropy than that of $\lambda$.}
\label{fig:lambda-theta-entropy}
\vspace{-0.5cm}
\end{figure}

\bpara{Overview of Packet Out-of-Order Delivery.}
We first present a brief overview of packet out-of-order delivery under different parameter settings. 
We use reorder entropy (referred to as \emph{entropy} hereafter for the ease of expression) to quantify the level of packet out-of-order.
By tuning the values for the parameters, namely $\theta$ and $\lambda$, we make observations on how the entropy changes accordingly. 
The experimental results are depicted in Figure~\ref{fig:lambda-theta-entropy}. 
As we can observe that, the entropy declines along the $\theta$ axis, e.g., when $\theta$ increases, the threshold for PABO to bounce back packets becomes higher and thus, less packets will be bounced under a certain traffic condition, as a result of which packet out-of-order is less severe. Along the $\lambda$ axis, the variation of entropy is less significant. Note that the entropy converges to a minimal point in cases that the switch will only bounce back packets when the buffer is full.


\bpara{Separate Impact of Parameters $\theta$ and $\lambda$.}
We now focus on analyzing how $\theta$ and $\lambda$ affect on packet out-of-order separately.
In addition to the entropy which measures the level of packet out-of-order,  we calculate the \emph{variance} of the buffer utilization of all the switches in the network. Note that the variance can be used to roughly characterize the effectiveness of PABO as PABO utilizes the buffer of upstream switches to avoid packet loss, leading to a more even distribution of packets among the switch buffers in the network. Meanwhile, we define \emph{timeRatio} to measure the scope of affected packets under different pairs of $\theta$ and $\lambda$. For those utilization of the buffers that will be considered during the probabilistic decision making process, we calculate the average time ratio of them over $\theta$.

\begin{figure}
\begin{minipage}{0.3\textwidth}
\centerline{\includegraphics[width=5.7cm]{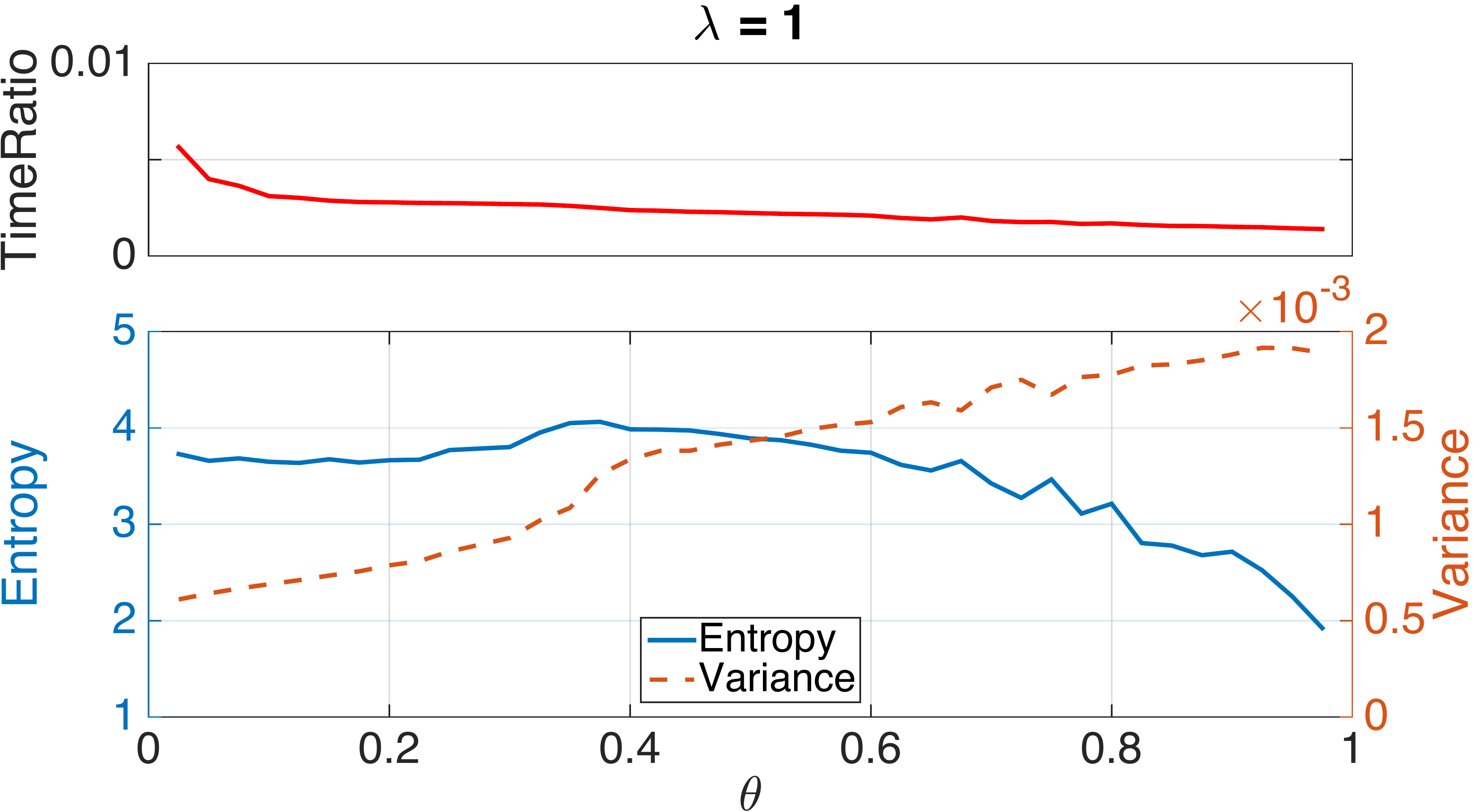}}
\centerline{(a)}
\end{minipage}
\hfill
\begin{minipage}{0.3\textwidth}
\centerline{\includegraphics[width=5.7cm]{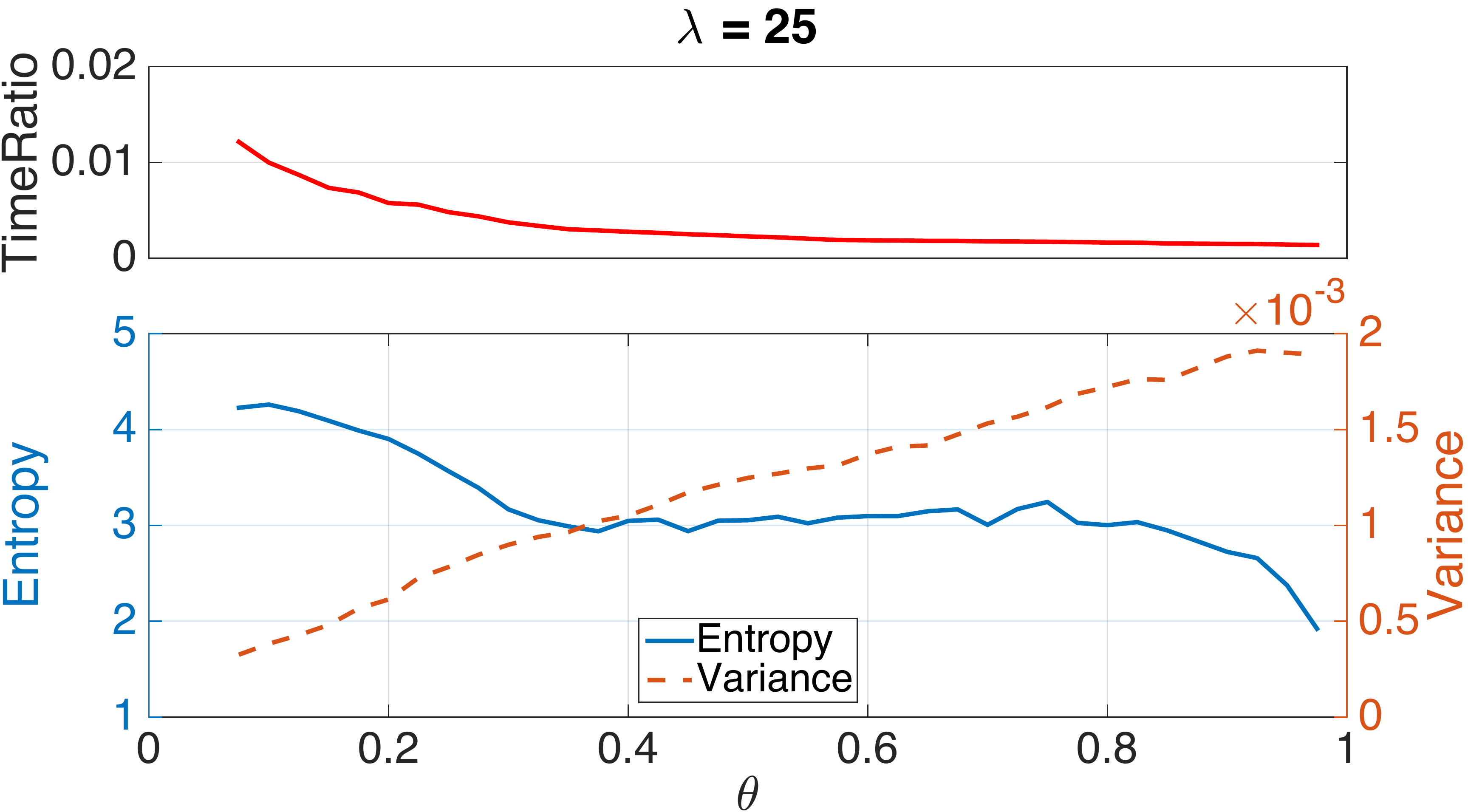}}
\centerline{(b)}
\end{minipage}
\hfill
\begin{minipage}{0.3\textwidth}
\centerline{\includegraphics[width=5.7cm]{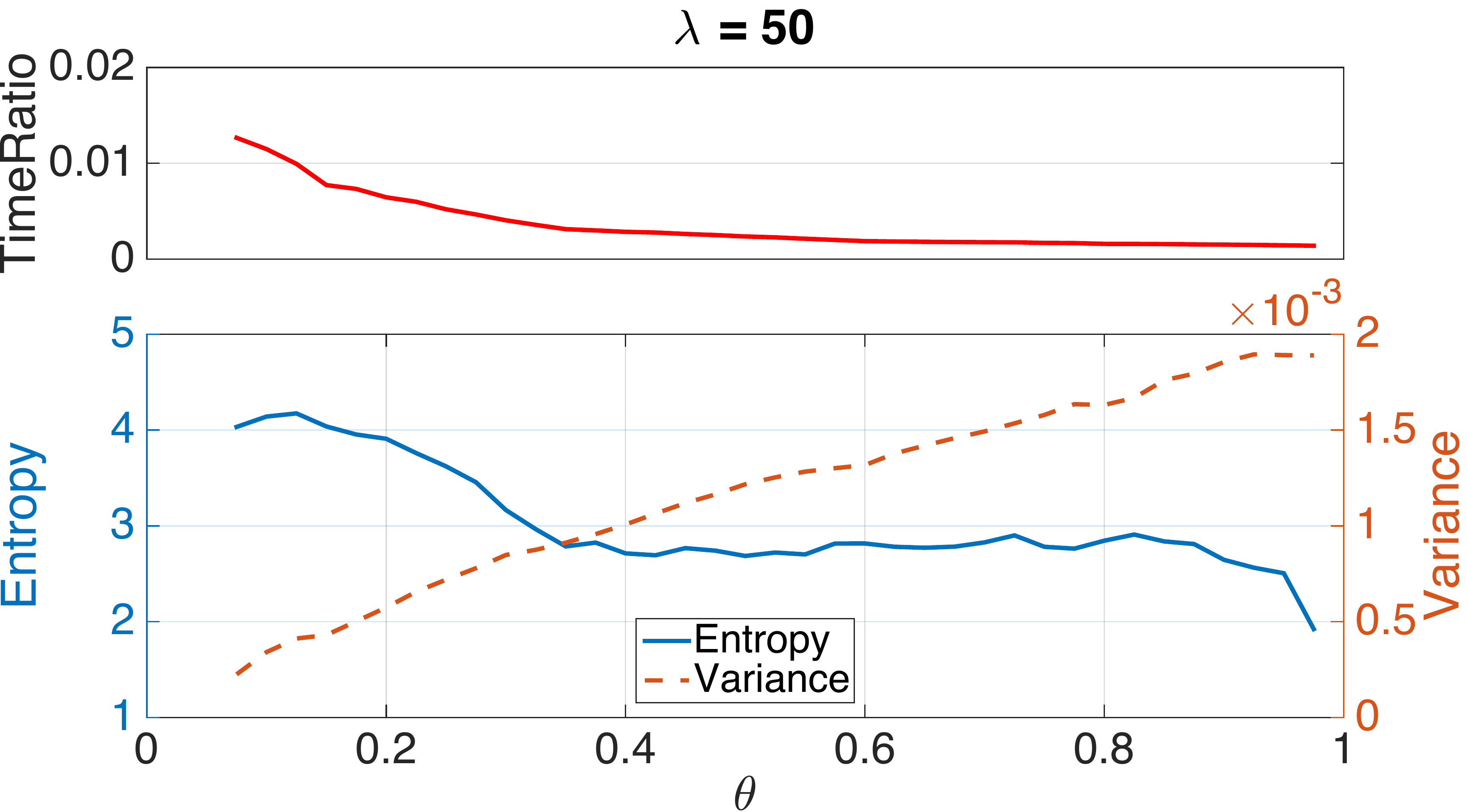}}
\centerline{(c)}
\end{minipage}
\vfill

\begin{minipage}{0.3\textwidth}
\centerline{\includegraphics[width=5.7cm]{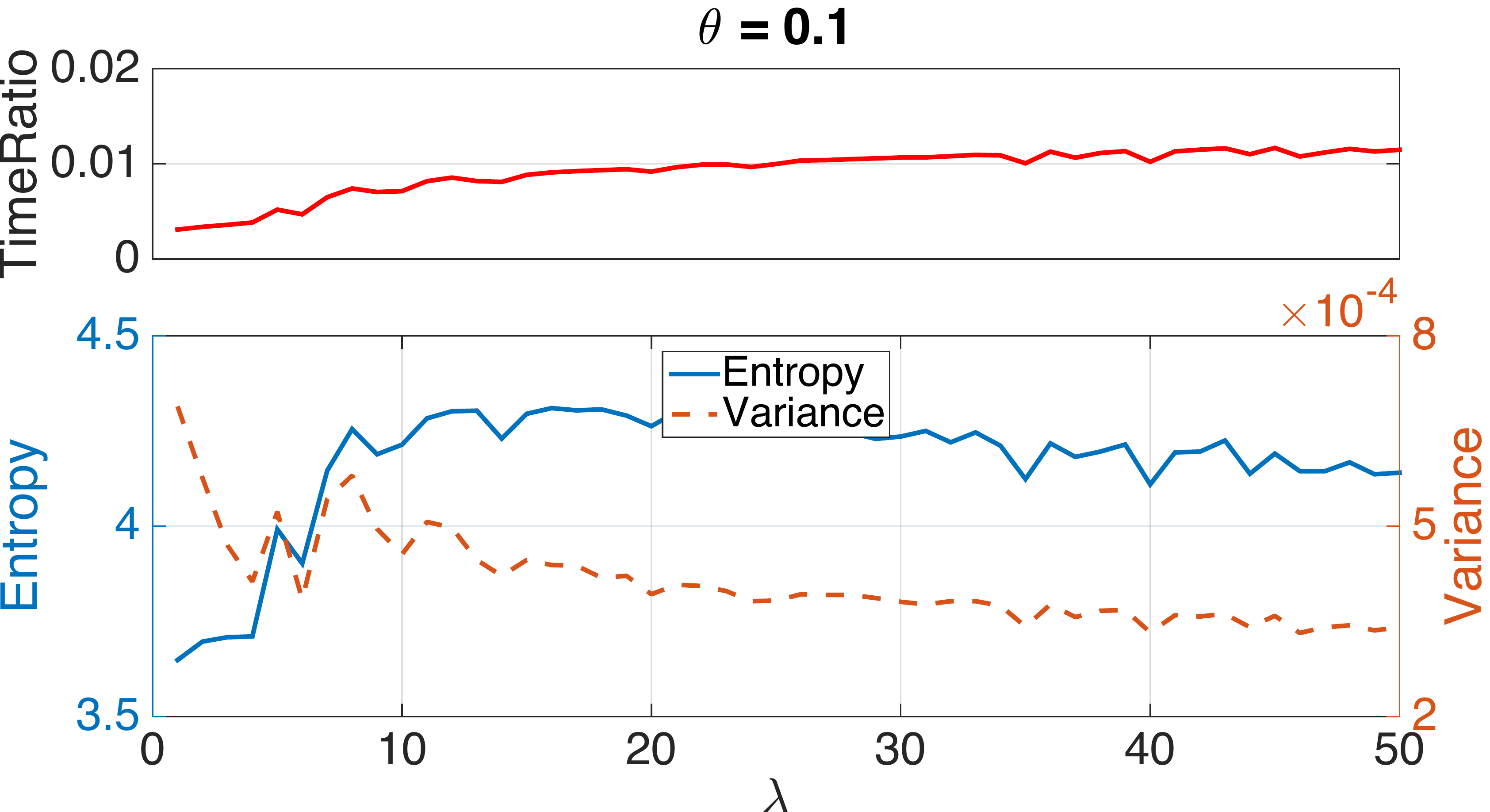}}
\centerline{(d)}
\end{minipage}
\hfill
\begin{minipage}{0.3\textwidth}
\centerline{\includegraphics[width=5.7cm]{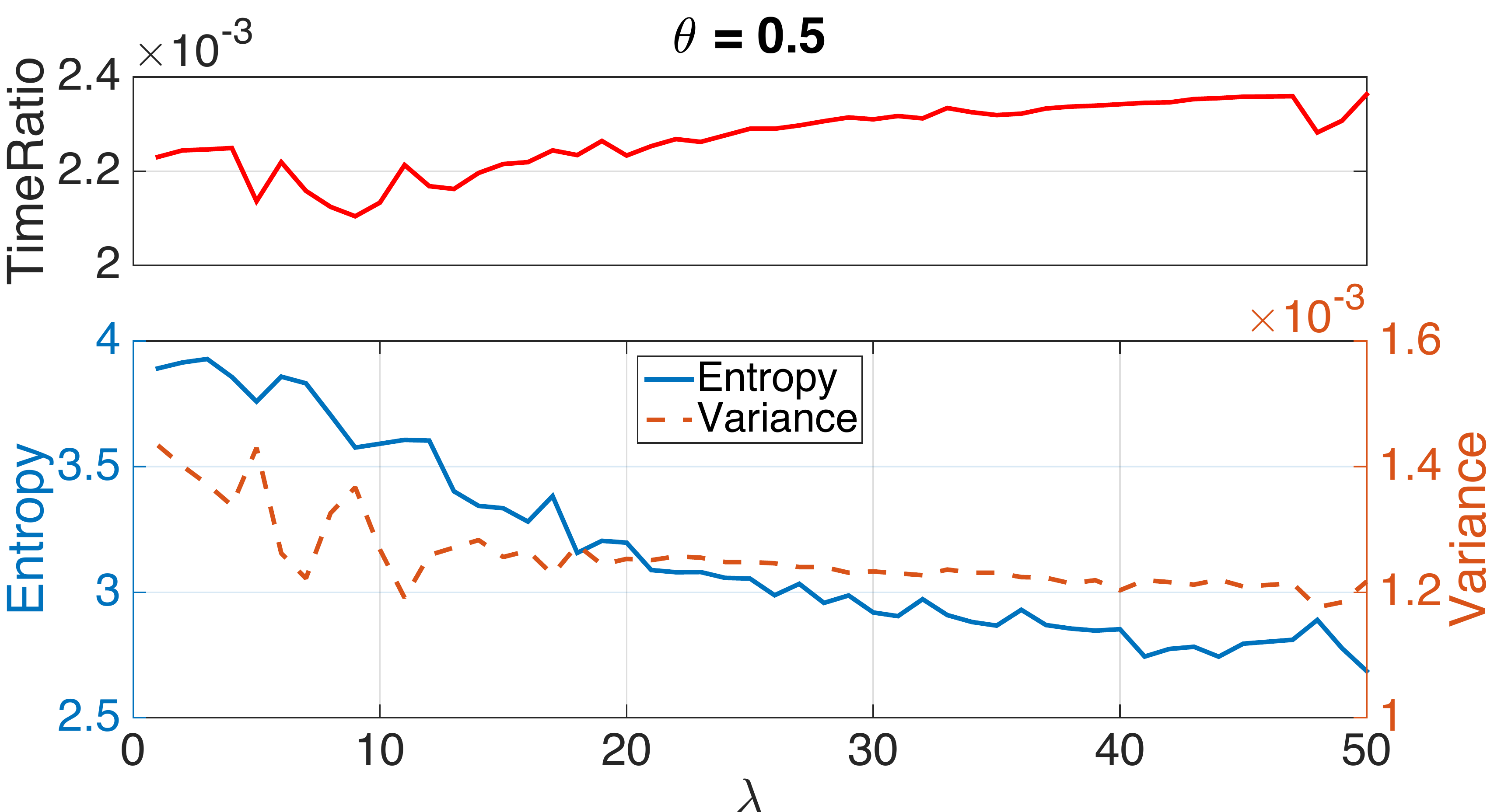}}
\centerline{(e)}
\end{minipage}
\hfill
\begin{minipage}{0.3\textwidth}
\centerline{\includegraphics[width=5.7cm]{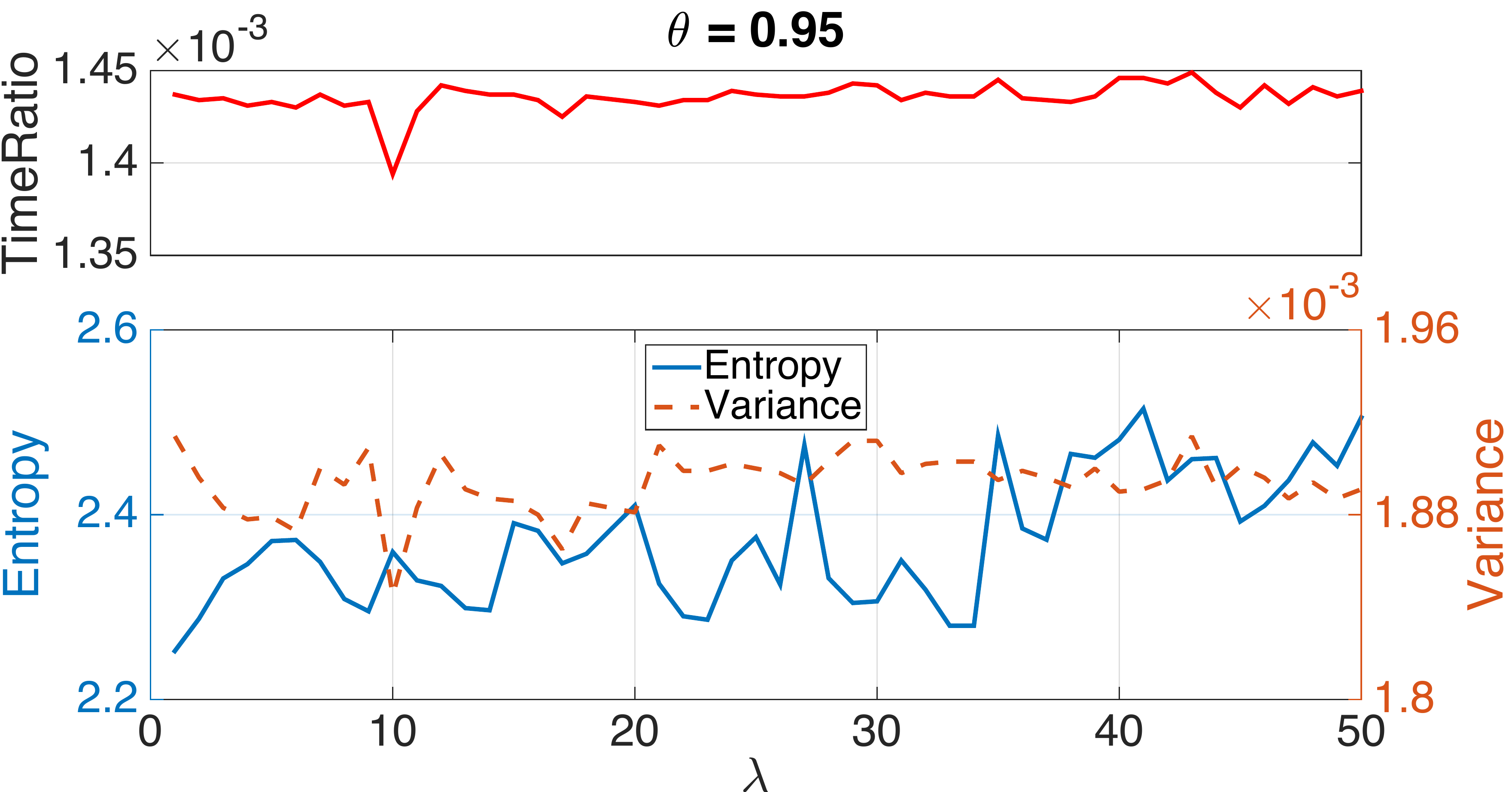}}
\centerline{(f)}
\end{minipage}
\caption{Separate impact of parameter $\theta$, $\lambda$ in bounce probability function on  entropy, variance and timeRatio: (a)-(c) the influence of fixed $\lambda$ and  change of $\theta$, and (d)-(f) the influence of fixed $\theta$ and  change of $\lambda$.}
\label{fig:entropy-variance-timeRatio} 
\vspace{-0.5cm}
\end{figure}

Figure~\ref{fig:entropy-variance-timeRatio} depicts the separate impact of $\theta$ and $\lambda$ on the  entropy, the variance, and the timeRatio.
We select three representative values for $\theta$ and $\lambda$ respectively to conduct further analysis. 
It can be generated observed that the entropy in all the six figures is highly correlated to timeRatio, which confirms that PABO can affect the level of packet out-of-order.

Figure~\ref{fig:entropy-variance-timeRatio} (a)-(c) illustrate the impact of parameter $\theta$ on packet out-of-order. According to Equation~(\ref{eq:prob_func}), larger $\theta$ means the less effectiveness of PABO. We can observe in the three figures that timeRatio reflecting the affected scope of PABO decreases gradually to a steady point. This is because most switches are under low utilization, thus the change of $\theta$ in a lower range can affect more switches.
As PABO is gradually losing its influence, we can observe that the variance in the three figures grows with similar tendency. 
However, the decrease tendency is different for the entropy.
In Figure~\ref{fig:entropy-variance-timeRatio}(a), the entropy changes relatively stable comparing to the other two figures. This can be explained by the change of affected scope of packets: when $\lambda$ is at small value (e.g. $\lambda=1$), the change of $\theta$ has much less impact on timeRatio. In Figure~\ref{fig:entropy-variance-timeRatio}(b) and (c), the entropy drops sharply at first, then remains stable in the middle area, followed by a further decrease in the end. This is in accordance with the change of timeRatio. The final decline is reasonable, as the switches are trying to avoid packet bounces as much as possible.



Figure~\ref{fig:entropy-variance-timeRatio} (d)-(f) depict the influence of parameter $\lambda$ on packet out-of-order. It is known that larger $\lambda$ means the more effectiveness of PABO.
In pace with the growth of $\lambda$, timeRatio increases and the variance decreases with a reasonable range of fluctuation.
Moreover, we can observe from the ordinate range of the above figures that larger $\theta$ leads to the reduced influence of $\lambda$ on timeRatio. When $\theta$ is very close to 1 (e.g. in Figure~\ref{fig:entropy-variance-timeRatio} (f)), the influence of $\lambda$ can be ignored, since it hardly affects the bounce probability.
As is demonstrated in Figure~\ref{fig:entropy-variance-timeRatio} (d) that the entropy rises rapidly at first and then remains a steady point. However, Figure~\ref{fig:entropy-variance-timeRatio} (e) illustrates an opposite tendency of entropy. This is because when $\theta=0.5$, S7 is the only switch among the topology in Figure~\ref{fig:exp-topo} that satisfies the condition for bouncing back packets. As $\lambda$ rises, the percentage of bounced packets continues to grow until it is infinitely close to one. In this case, higher bounce back percentage at S7 leads to lower entropy, as the bouncing back process of the packets is comparable with experiencing an equally extended path.

\subsection{Discussion}
As stated above, the effectiveness of PABO can be a tradeoff when deciding the values of  $\theta$ and $\lambda$.
The main idea of PABO is to utilize the buffer of upstream switches to relieve traffic burdens on the bottleneck switch,  making the packets more evenly distributed in the network. Consequently,  the transient congestion (e.g. the incast problem) is relieved.
Higher influence of PABO brings the stronger ability to mitigate transient congestion in the network, which can also result in more severe packet out-of-order.
Severe packet out-of-order can lead to the performance collapse of the related TCP connection. Therefore, we should balance between variance (representing the effectiveness of PABO) and the entropy (reflecting the extent of out-of-order), by carefully choosing the parameters for target networks. 
Focusing on this tradeoff, the recommended solution is to control the bounce back scope around the congestion point. This means when there is a congestion, $\theta$ should be set to just exclude the switches which are not faced with the danger of packet loss.
There is no need to set the $\theta$ too large, for the reduced entropy is not worth to compensate for the increased variance. If the bounce back scope is limited to be just around the congestion point by $\theta$, larger $\lambda$ representing higher bounce back percentage is preferred to reduce both the entropy and the variance.

\section{Hop-by-Hop Evaluation}

We first conduct simulation studies at a hop-by-hop level to evaluate the performance of PABO, and we report the experimental results in this section.

\subsection{Simulation Setup}
We adopt the same topology as Section~\ref{sec:out-of-order} in Figure~\ref{fig:exp-topo}, except that the hosts don't contain any high layer protocols (i.e., IP, TCP).
We focus on only one direction data retransmission for the moment, where three senders (i.e., H1, H2 and H3) send data simultaneously to a single receiver (i.e., H4). The data from all the senders will be aggregated at the last-hop switch (i.e., S7), in which the congestion appears.
The duration of all the simulations is set to one second to cover multiple appearances of periodic congestions. The links in the network are assumed to have the same rate of 1Gbps.


\bpara{Traffic.}
We imitate a periodic uniform flow by altering the traffic-generating module. 
We fix the burst number to be one and set the $\mathtt{sendInterval}$ to small values. Then, we introduce a new parameter $\mathtt{numPacketsPerGenerate}$ to specify the number of packets to be sent in each generating. The duration of each generating can be roughly calculated by $\mathtt{numPacketsPerGenerate}$ $\times$ $\mathtt{sendInterval}$. As the overlap of two consecutive generatings may bring persistent congestions which will overflow the sender's buffer, we modify the $\mathtt{EtherTrafGen}$ module to enable \emph{pause} after every generating. 
To guarantee fine-grained control over the sending rate, we fix the value of $\mathtt{sendInterval}$ to be 10$\mu$s. Then, we control the rate by tuning $\mathtt{numPacketsPerGenerate}$ to simulate congestions in different severities. We set $\mathtt{pauseInterval}$ to be 0.2s, which is sufficient to avoid overlaps between two consecutive congestions.

\bpara{Queue.}
The input and output queues of both hosts and switches are with the $\mathtt{DropTailQueue}$ type. When using PABO, the capacities of both $\mathtt{normalQueue}$ and $\mathtt{bounceQueue}$ in switches are set to be 500 by default. And we specially allocate larger buffers of size 1500 to both $\mathtt{normalQueue}$ and $\mathtt{bounceQueue}$ in senders to avoid packet loss at the sender side.
In the cases without PABO where $\mathtt{bounceQueue}$ are not used, we double the capacities of $\mathtt{normalQueue}$ in both switches and senders for fairness concern.

\bpara{Packets.}
All the packets generated are in the IEEE 802.3 frame formats with the payload size set to be 1500 bytes.

\subsection{Effectiveness under Different Congestion Severities}
We validate the effectiveness of PABO by comparing it to the standard link-layer protocol under three different severities of congestion. Parameters of bounce probability function $P$ are fixed as $\lambda = 50$, $\theta = 0.8$. We set $\mathtt{numPacketsPerGenerate}$ to 500, 1500, 2500 to simulate different severities of congestion, which are respectively referred to as \textit{mild}, \textit{moderate}, and \textit{severe}. We also measure the cases without PABO under the same traffic conditions and experimental results show a packet drop rate of $0.13\%$, $44.46\%$, $53.34\%$ at S7, respectively. Note that \emph{the retransmission of those lost packets by upper-layer protocols could generally result in an order of magnitude increase on packet delay due to the timeout-based fashion \cite{Chen-Incast-2009}.}

\begin{figure}[!t]
\centering
\subfigure[]{
\label{fig:pabo_perf}
\includegraphics[scale=0.3]{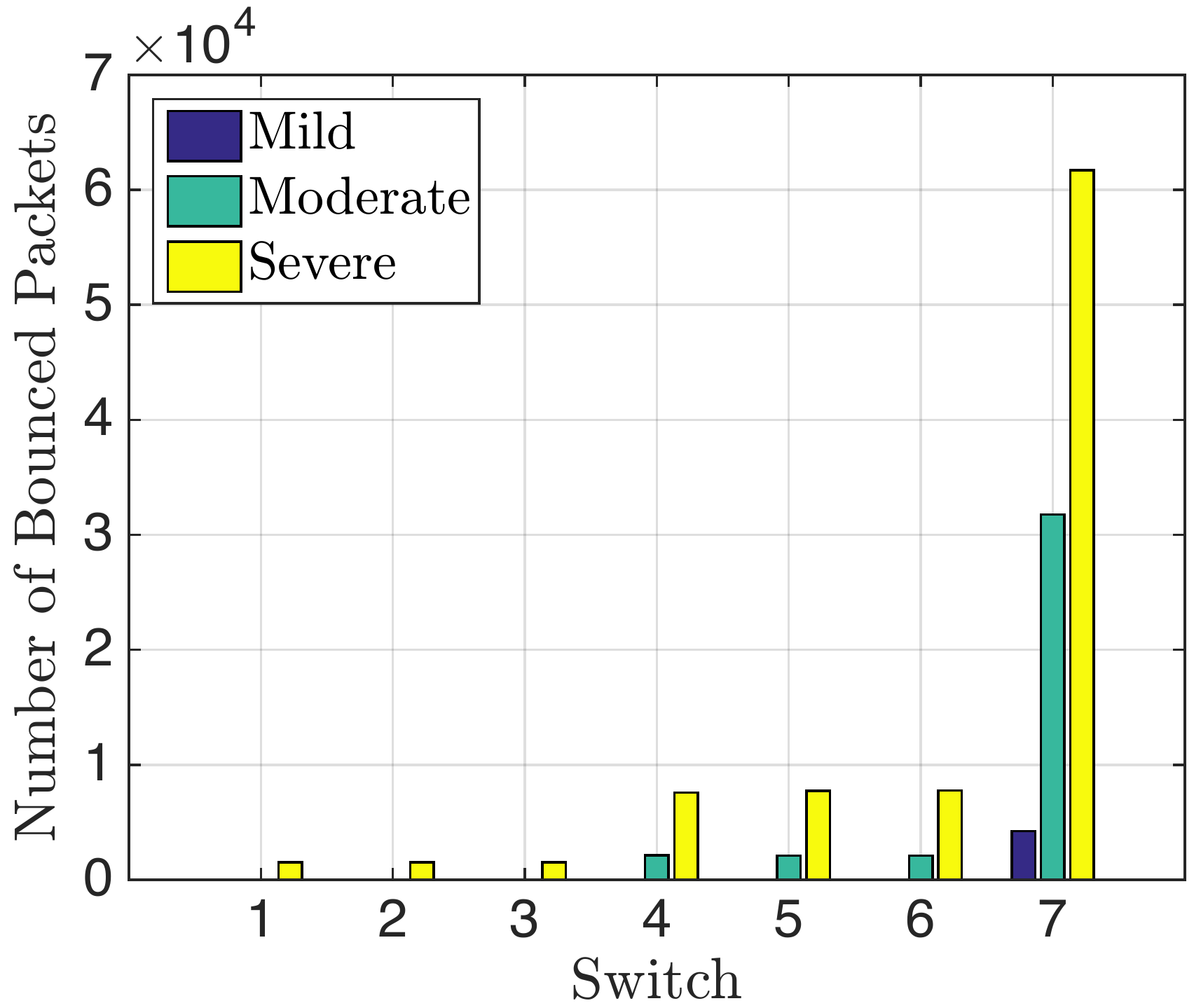}
}
\hspace{-0.2cm}
\subfigure[]{
\label{fig:PABO-TotalHopNumDistributionCDF}
\includegraphics[scale=0.3]{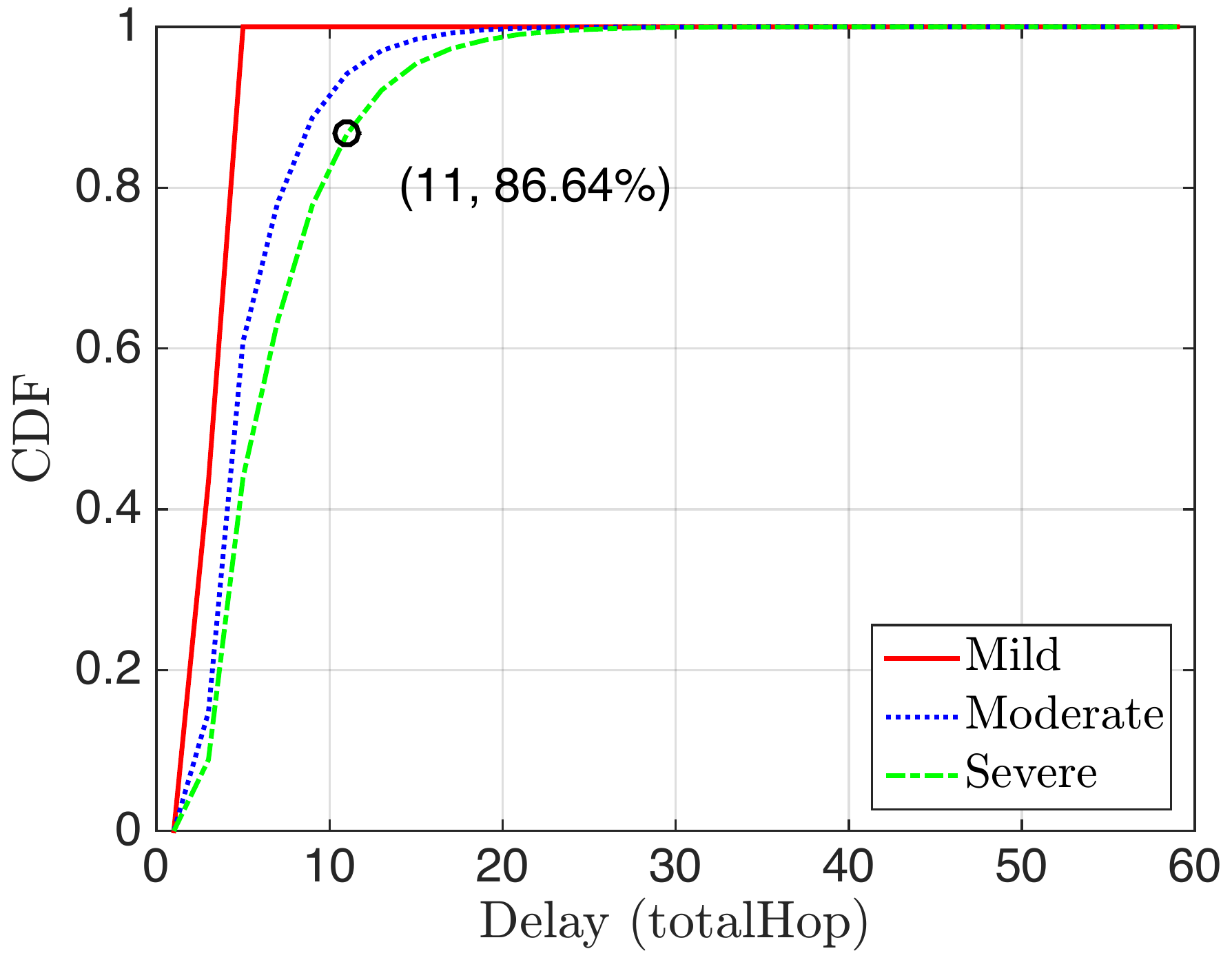}
}
\vspace{-0.3cm}
\caption{Performance results of PABO: (a) Packet bounce frequency under different severities of congestion, and (b) the CDF of the packet delay measured by $\mathtt{totalHop}$.}
\label{fig:PABO_Performance} 
\vspace{-0.5cm}
\end{figure}

When PABO is involved, packet loss can be prevented. The number of bounced packets at each switch in all the three scenarios is illustrated in Figure~\ref{fig:pabo_perf} and the proportion of $\mathtt{maxBouncedDistance}$ is shown in Table~\ref{tab:PassBackDistanceDistribution}. As we can observe that, only switch S7 has bounced $56.55\%$ of all the packets in the mild scenario. When the extent of the congestion becomes larger, as in the moderate scenario, switches that are one hop from the receiver (i.e. H4), have bounced packets and there are in total $85.27\%$ of the packets have been bounced one hop away from its last normally reached switch. When the congestion becomes very severe, all the switches will be activated for bouncing packets, while there are still up to $8.82\%$ of the packets being successfully transmitted without any interference.

\begin{table}[!t]
\centering
\caption{Distribution of $\mathtt{maxBouncedDistance}$}
\vspace{-0.3cm}
\label{tab:PassBackDistanceDistribution}
\begin{tabular}{ r | c | c | c | c }
\hline
\multirow{2}{*}{\textbf{Scenario}} & \multicolumn{4}{c }{\textbf{\# of $\mathtt{maxBouncedDistance}$}} \\ \cline{2-5}
                 &  \textbf{0}   &  \textbf{1}   &  \textbf{2}   &  \textbf{3}  \\ \hline \hline
Mild & 43.45\% & 56.55\% & -- & --\\ \hline
Moderate & 14.73\% & 85.27\% & -- & --\\ \hline
Severe & 8.82\% & 84.14\% & 7.04\% & --\\ \hline
\end{tabular}
\vspace{-0.3cm}
\end{table}

The zero packet loss guarantee is achieved at the sacrifice of delay, as bouncing a packet would inevitably increases its $\mathtt{totalHop}$. To measure the delay stretch brought by PABO, we collect the values for $\mathtt{totalHop}$ from all the packets in three scenarios and presented the CDFs in Figure~\ref{fig:PABO-TotalHopNumDistributionCDF}. We can observe that almost all packets experience a delay no more than $5$ hops in the mild scenario, and up to $86.64\%$ of the packets traverse no more than $11$ hops in the severe scenario, which are still no more than four times the delay in the normal case. This is quite acceptable compared to the orders of magnitude delay increases in retransmission-based approaches.

\begin{figure}[!t]
\centering
\subfigure[]{
\label{fig:PABO-DataQueueUtilizationOfSwitchAll}
\includegraphics[scale=0.3]{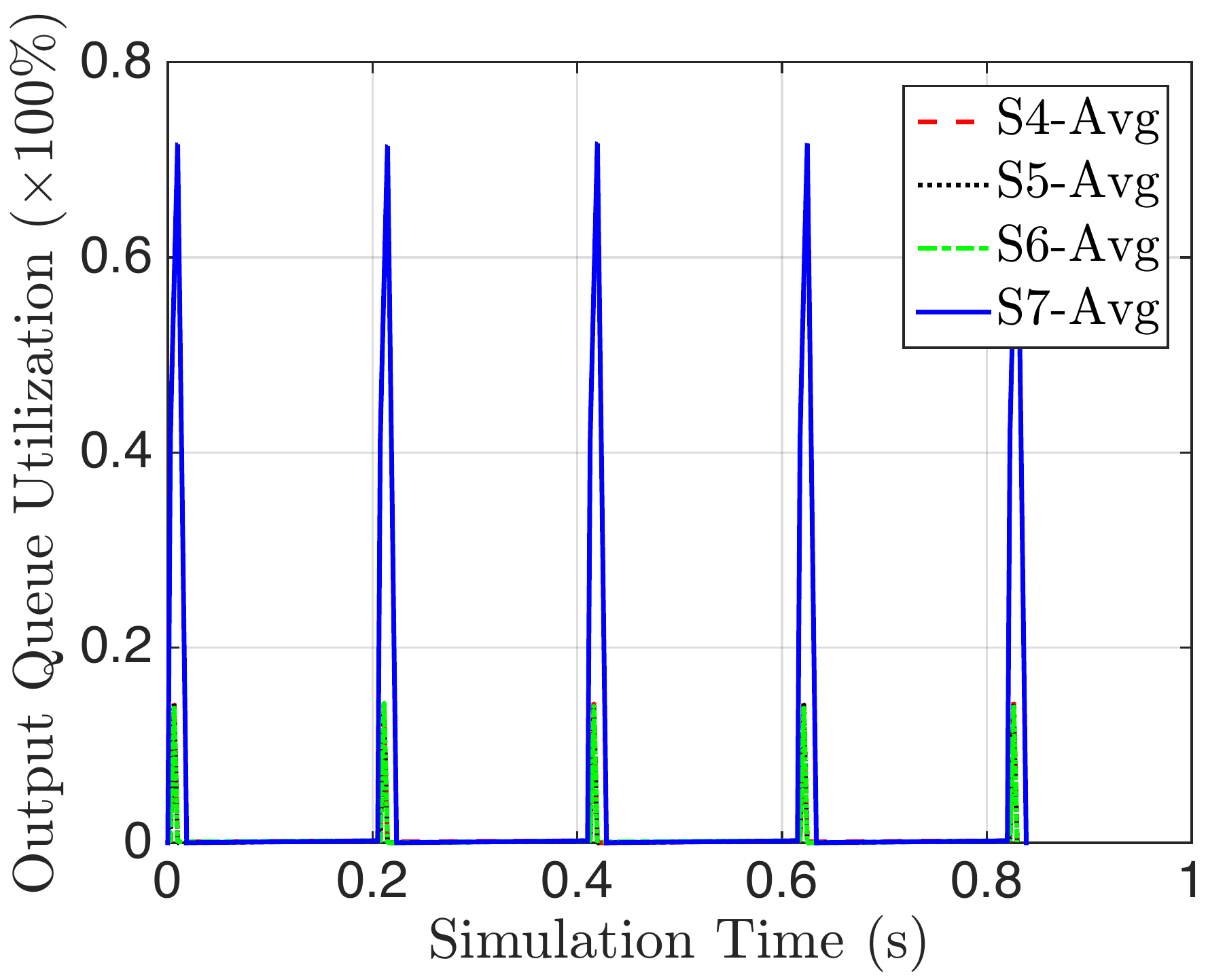}
}
\hspace{-0.15cm}
\subfigure[]{
\label{fig:PABO-DataQueueUtilizationOfSwitchOne}
\includegraphics[scale=0.3]{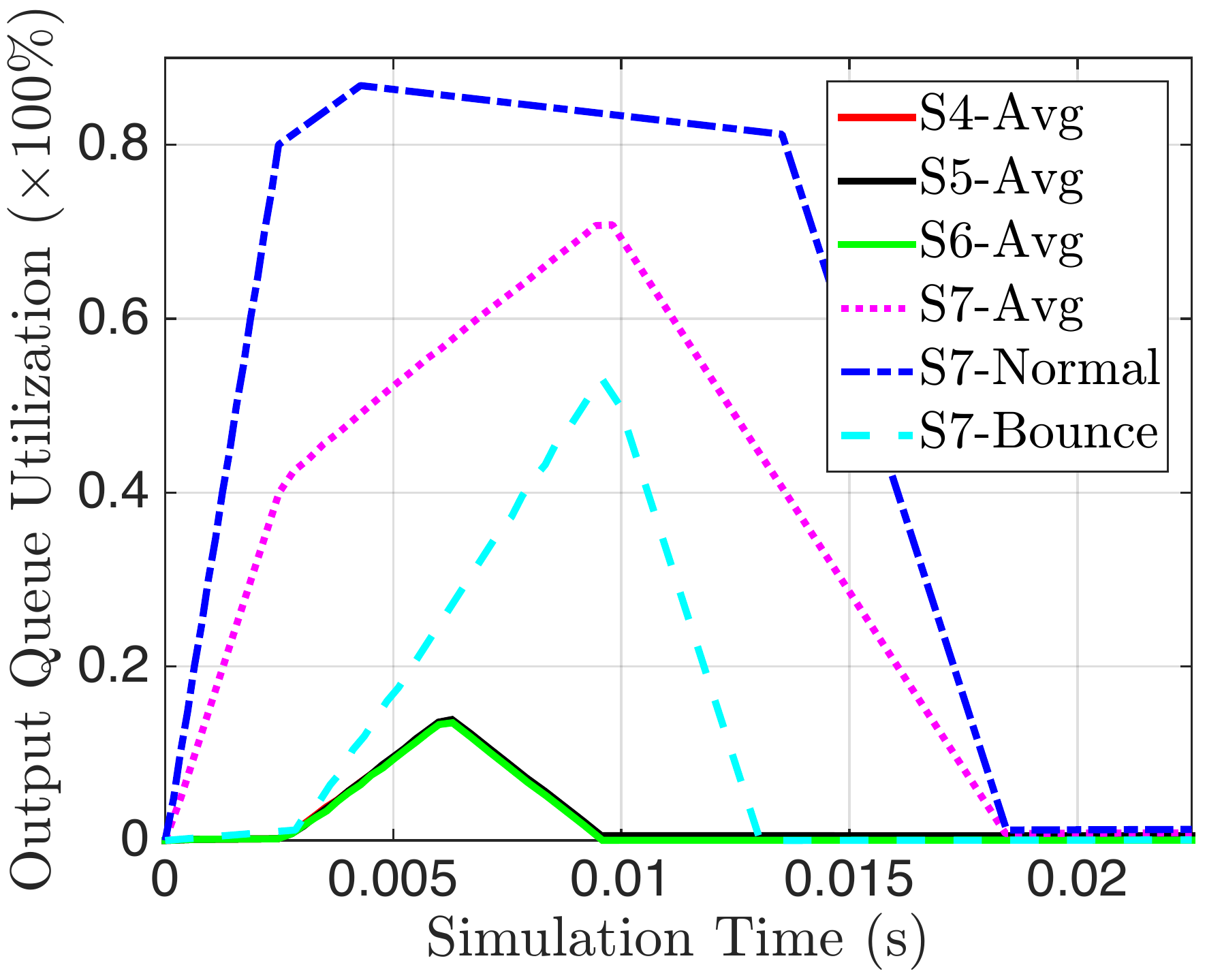}
}
\vspace{-0.3cm}
\caption{Switch output queue utilization under the mild scenario: (a) over the whole duration of the simulation, and (b) over the first traffic peak.}
\label{fig:PABO-DataQueueUtilizationOfSwitch} 
\vspace{-0.7cm}
\end{figure}

We also monitor the output queue utilization of switch S4, S5, S6, and S7 in the mild scenario, and the results are depicted in Figure~\ref{fig:PABO-DataQueueUtilizationOfSwitch}. We monitor the utilization levels of $\mathtt{normalQueue}$ and $\mathtt{bounceQueue}$ separately, and calculate the average utilization of the two queues at the output. Figure~\ref{fig:PABO-DataQueueUtilizationOfSwitchAll} depicts the average utilization of the relevant switches over the whole duration of the simulation which includes five appearances of transient congestion. We then focus on the first traffic peak as illustrated in Figure~\ref{fig:PABO-DataQueueUtilizationOfSwitchOne}, where we notice that when the $\mathtt{normalQueue}$ utilization of S7 becomes high, packets are bounced to upstream switches S4, S5, and S6 and thus, the $\mathtt{bounceQueue}$ at S4, S5, S6, as well as S7, will be used instead of overflowing the $\mathtt{normalQueue}$ of S7. When the traffic volume declines, the queues at S4, S5, and S6 will be firstly cleared up and then finally the congestion vanishes with the drop of the average (first bounce and then normal) queue utilization of S7. This verifies our claim that PABO can avoid packet loss and handle congestion by temporarily utilizing the buffers of upstream switches.

\begin{figure}
\centering
\subfigure[]{
\label{fig:PABO_ThetaResults}
\includegraphics[scale=0.3]{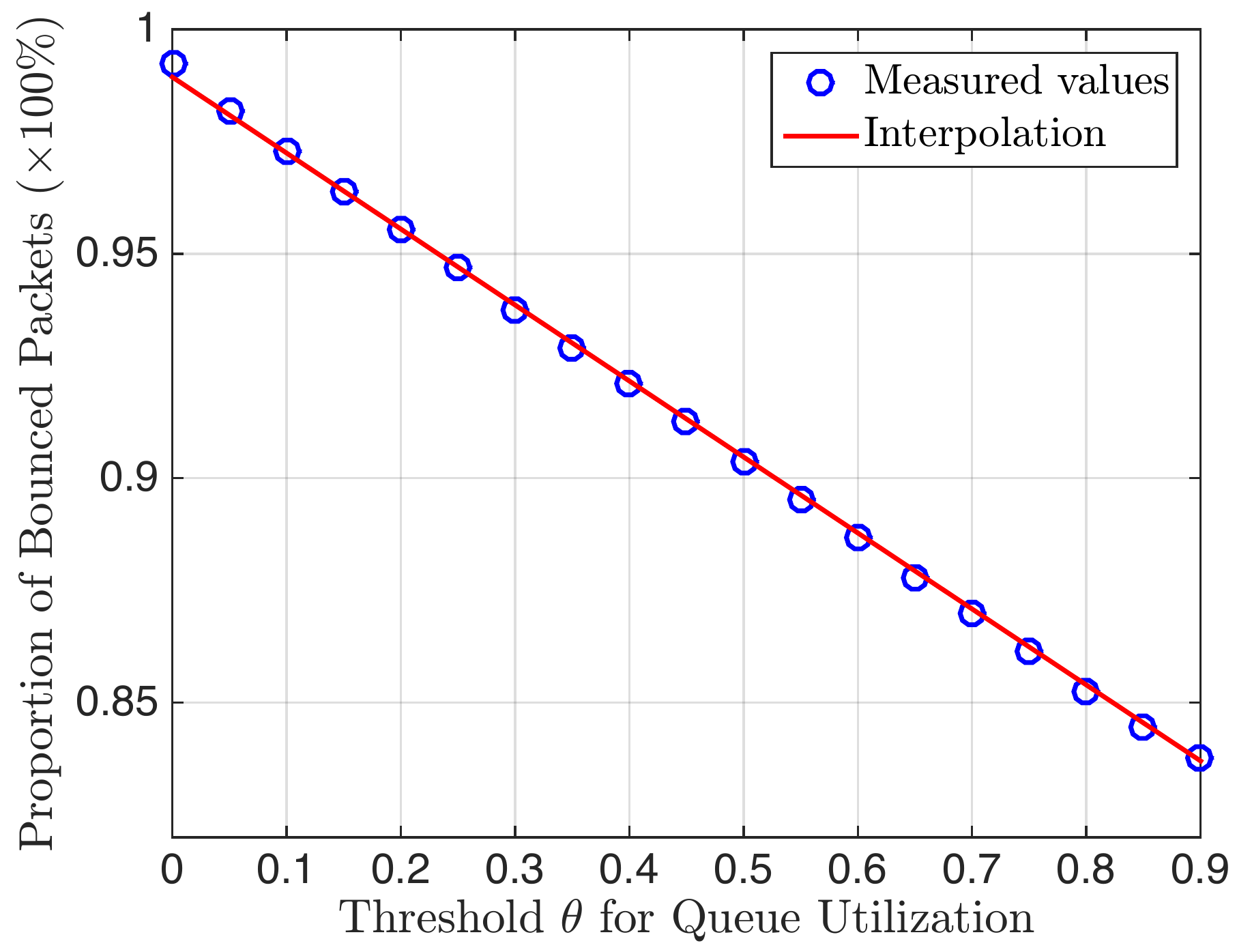}
}
\hspace{-0.2cm}
\subfigure[]{
\label{fig:PABO_Lambda_Results2}
\includegraphics[scale=0.3]{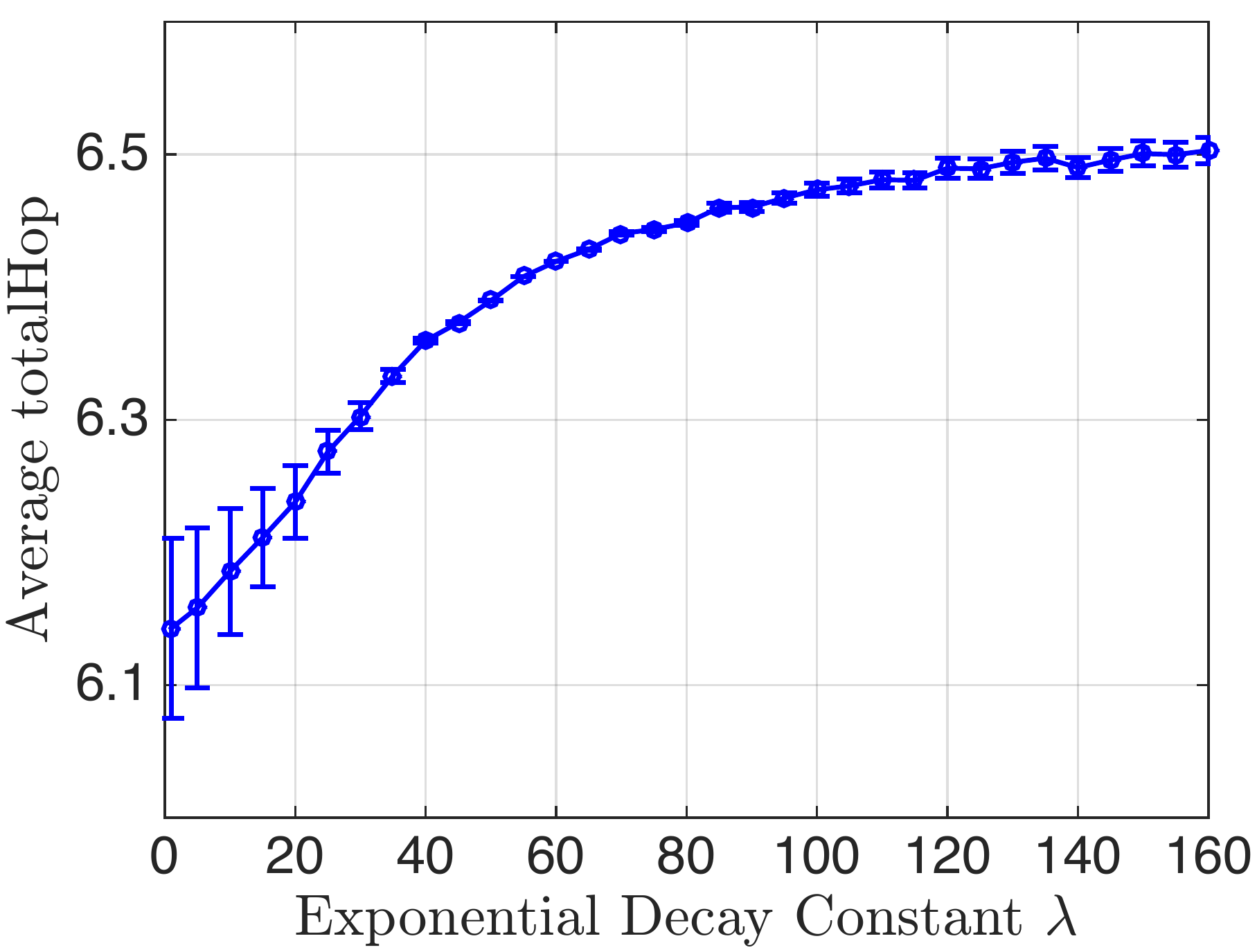}
}
\vspace{-0.3cm}
\caption{Impact of the parameters (a) $\theta$ and (b) $\lambda$ on the proportion of bounced packets and the average packet delay measured by $\mathtt{totalHop}$, respectively. The proportion of bounced packets decreases linearly with the increase of $\theta$, as $\theta$ is the threshold for PABO begins to work.  The average $\mathtt{totalHop}$ of each packet increases gradually stable with the increase of $\lambda$, for the growth of $\lambda$ means the increasing effectiveness of PABO.}
\label{fig:PABO_Parameters} 
\vspace{-0.6cm}
\end{figure}%

\subsection{Impact of Parameters}

We also explore the impact of the parameters on the effectiveness of PABO in the moderate scenario. We focus mainly on two parameters in the bounce probability function $P$ and the experimental results are shown in Figure~\ref{fig:PABO_Parameters}. We measure the impact of threshold $\theta$ for queue utilization on the proportion of packets being bounced and the exponential decay constant $\lambda$ on the average number of total hops for all packets, respectively. When setting $\lambda$ to a fixed value $50$, we notice a clear trend that the proportion of bounced packets decreases linearly with the increase of $\theta$, as depicted in Figure~\ref{fig:PABO_ThetaResults}. Similarly, we fix $\theta$ to be $0.8$ and observe that the average number of total hops, i.e., $\mathtt{totalHop}$, increases gradually stable with the increase of $\lambda$ from $0$ to $160$. Thereafter, it remains stable with only a negligible variation.

The values for the parameters should be determined according to the needs of the network operator. The general principle is: smaller $\theta$ and larger $\lambda$ improve the effectiveness of PABO, which tends to avoid the congestion at a earlier time; while larger  $\theta$ and smaller $\lambda$ would prefer to reduce the sensitivity of PABO, thus delay the absorption of the congestion.

\section{End-to-End Evaluation in Realistic Topology}

We further evaluate the performance of PABO in the Fattree network topology \cite{Al-Fattree-2008} and we present the experimental results in this section.
We simulate different traffic patterns (many-to-one, many-to-many) to observe the performance of PABO, then we present the parameter study of  $\theta$ about the impact of PABO on entropy, per-packet delay and flow completion time.

\subsection{Fattree Implementation}
\label{sec:routing}
We implement the Fattree routing algorithm (used in two-level routing table) on $\mathtt{EtherSwitch}$, and calling it the $\mathtt{FattreeSwitch}$. For the reason that this module does not contain network layer, we implement the two-level routing table on the link layer by applying the following modifications.

\bpara{Addressing.}
Firstly, in line with the Fattree addressing scheme, we make a one-to-one mapping between an IP address and a MAC address. To illustrate, the IP address 10.0.0.1 corresponds to the MAC address 0A-AA-0A-00-00-01. All MAC addresses share the same first two octets 0A-AA, and the rest are transformed equivalently.
Then we assign the transformed MAC addresses to each switch and host. Note that all the MAC ports of a switch share the same MAC address for simplicity, which will not effect any experimental results.

\bpara{Structure.}
Secondly, we modify the structure of $\mathtt{MACTable}$ to allow entries containing prefixes and suffixes (i.e., /m prefixes are the masks used for left-handed matching, /m suffixes are the masks used for right-handed matching).

\bpara{Lookup.}
Thirdly, we modify the lookup unit of $\mathtt{MACTable}$ to allow two-level route lookup.
Prefixes are intended for route matching of intra-pod traffic, while suffixes for inter-pod traffic.
The value of prefix or suffix is simply used to check the number of octets required for comparison. To be more specific, if we want to match an entry in the $\mathtt{MACTable}$ with a left-handed prefix of N (e.g. $24$), we should find from left to right at least N/8 (e.g. 3) identical octets between this entry and the destination MAC address, excluding the same first two octets. This also applies to the match of a right-handed suffix except that the matching direction is from right to left.

\bpara{Routing Example.}
Here we explain the Fattree two-level routing algorithm implemented in Figure~\ref{fig:fattree-topo}.
For the hosts connected to each lower-level switch (e.g. S1) in this Fattree structure, the last octets of the left hosts are 02, and the last octets of the right hosts are 03.
Based on the last octet of the destination MAC address, the algorithm uses the prefix/suffix matching to disperse different traffic, which we will explain in a simplified way.
In the Fattree topology, each pod follows similar rules on packet routing.
We take pod 0 as an example to explain inter-pod routing and intra-pod routing separately.
For ease of expression, we refer to S1, S3 as the left-side switches, and S2, S4 as the right-side switches.
For inter-pod traffic, the left-side switches route a packet destined for another pod to the port number same as the last octet of its destination MAC address (e.g. packet addressed to 02 forwarded to port 2 and  03  forwarded to port 3).
The right-side switches work in an opposite way (e.g. packet addressed to 02  forwarded  to port 3 and  03  forwarded  to port 2).
We give an example to explain the route decisions taken for a packet from the inter-pod traffic: source H2 to destination H10, which is illustrated in Figure~\ref{fig:fattree-topo}.
Marked in the picture are the port numbers of the switches.
As the last octet of its destination is 03, the packet first take the port 3 of S1, then goes out at the port 2 of S4 to C3, after which there is only one path to take: to be transmitted to the destination pod, then the destination subnet switch where it is finally switched to its destination host.
For intra-pod traffic, the first-hop switch follows the same rules as in inter-pod routing, then the second-hop switch route the packet to its destination subnet switch, and finally the destination host.

\subsection{Simulation Setup}

\begin{figure}
\centering
\includegraphics[scale=0.6]{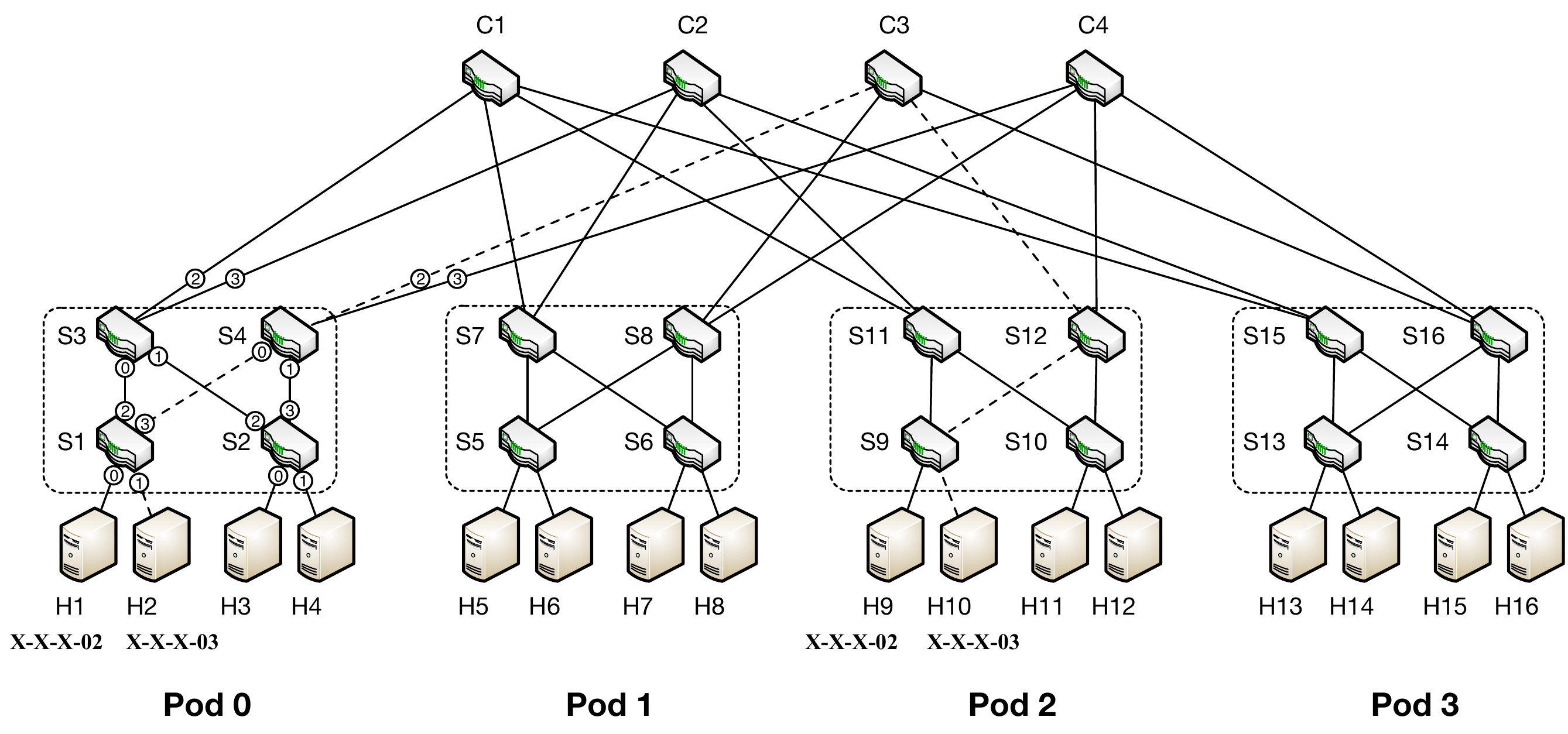}
\vspace{-0.3cm}
\caption{The Fattree topology used for evaluating the performance of PABO. Using the two-level routing tables described in Section~\ref{sec:routing}, packets from source H2 to destination H10 would take the dashed path.}
\label{fig:fattree-topo}
\vspace{-0.5cm}
\end{figure}

We use the Fattree network topology depicted in Figure~\ref{fig:fattree-topo} to evaluate the performance of PABO.
We initialize the $\mathtt{MACTable}$ of all the $\mathtt{FattreeSwitch}$ using input files that give all the prefixes and suffixes, and turn off the update function.
Meanwhile, to avoid the broadcast storm brought by ARP request of the hosts, we statically initialize all the hosts with the IP-address-to-MAC-address mapping information.
The duration of all the simulations is set to ten second, which can cover multiple appearances of periodic congestions. 
The links in the network are assumed to have the same rate of 1Gbps.

\bpara{Traffic.}
We simulate different traffic patterns by changing the number of servers and clients. For each TCP connection, we use the same traffic and buffer setup as Section~\ref{sec:out-of-order}, except that the advertised window is set to 50000 bytes.
When using PABO, we disable all the retransmissions (both fast retransmit and retransmission timeout) as well as skipping the related window reduction intended for congestion control, as the retransmissions are unnecessary due to the reliability of PABO. 
In the cases without PABO, we adopt the TCP Reno protocol to provide network congestion control.

\bpara{PABO Configuration.}
We set the system parameters $\theta= 0.95$, $\lambda=50$ in both many-to-one and many-to-many scenario experiments.

\subsection{Many-to-One Scenario}
First, we mimic the partition/aggregate traffic by specifying multiple servers responding to one client (i.e., H9) in a synchronized fashion. We simulate different severities of congestion by changing the number of servers (i.e., 3 to 1, 6 to 1, 9 to 1, 12 to 1). Then we observe how PABO performs in different congestions comparing to cases without PABO.

\begin{figure}
\centering
\includegraphics[scale=0.4]{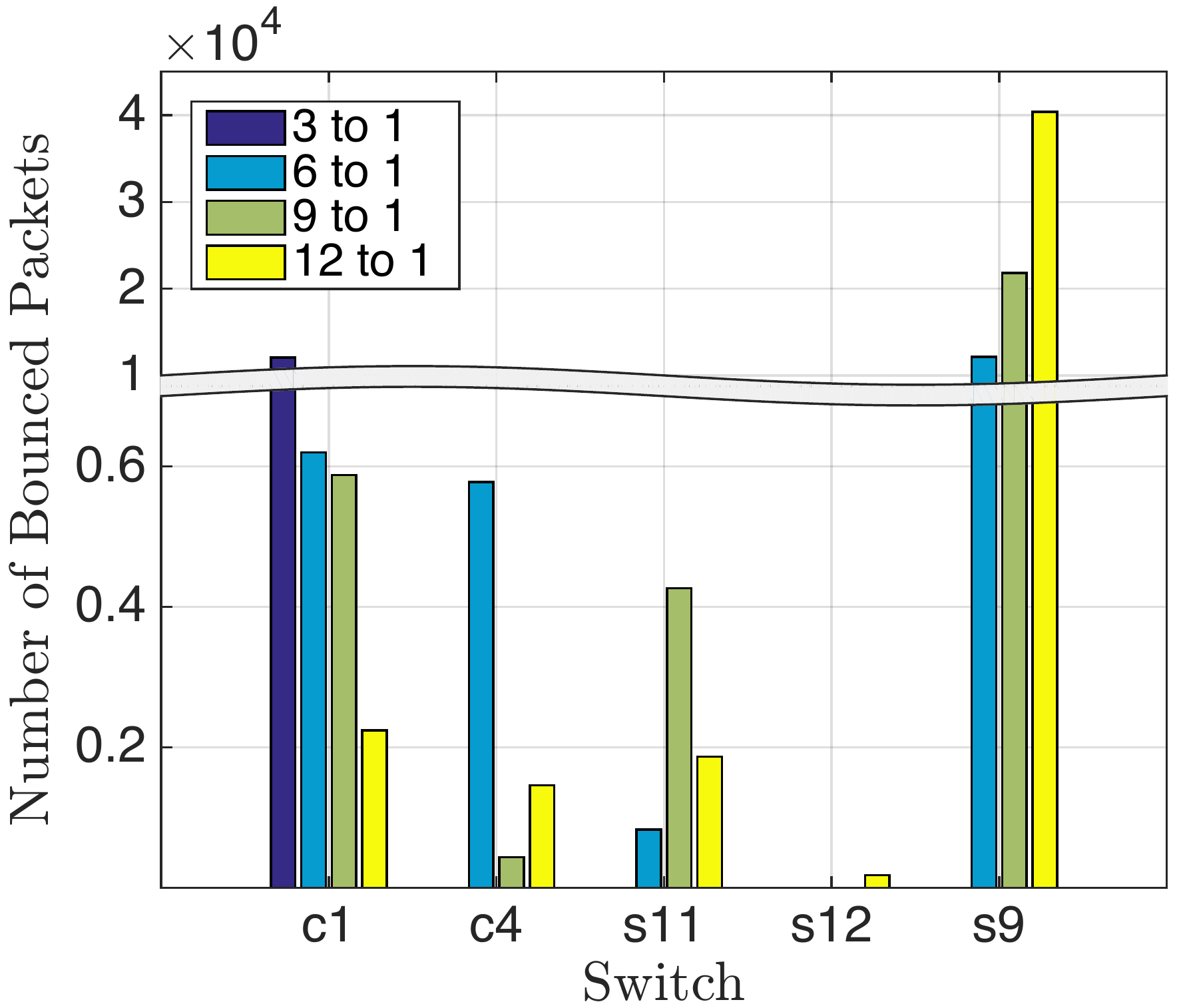}
\vspace{-0.3cm}
\caption{Packet bounce frequency of each switch under different many-to-one scenario, measured by number of bounced packets at each switch. The total frequency of a scenario increases with the growth of the number of servers.  The middle part is omitted using a break axis for the convenience of showing data with a wide range.}
\label{fig:pabo_perf_fattree}
\vspace{-0.5cm}
\end{figure}

\begin{table}[!t]
\centering
\caption{Distribution of $\mathtt{maxBouncedDistance}$}
\vspace{-0.3cm}
\label{tab:PassBackDistanceDistribution}
\begin{tabular}{ r | c | c | c }
\hline
\multirow{2}{*}{\textbf{Scenario}} & \multicolumn{3}{c }{\textbf{\# of $\mathtt{maxBouncedDistance}$}} \\ \cline{2-4}
                 &  \textbf{0}   &  \textbf{1}   &  \textbf{2}  \\ \hline \hline
3 to 1 & 50.17\% & 49.83\% & -- \\ \hline
6 to 1 & 53.24\% & 46.76\% & -- \\ \hline
9 to 1 & 56.71\% & 43.29\% & -- \\ \hline
12 to 1 & 57.68\% & 42.32\% & -- \\ \hline
m to m & 38.37\% & 61.63\% & -- \\ \hline

\end{tabular}
\vspace{-0.5cm}
\end{table}

Figure~\ref{fig:pabo_perf_fattree}  demonstrates the packet bounce frequency of each  switch under different many-to-one scenario.
In the 3 to 1 congestion scenario, we choose one host from each pod to be the servers except the pod with the client H9, i.e., H1, H5, H13. The first aggregate switch of the three connections is C1.
When PABO is not working, the core switch C1 will be the only switch to experience packet losses with an overall drop rate of 0.33\%.
As PABO participates, only  C1 bounce back packets to avoid packet loss at a bounce percentage of 49.83\%.

In the 6 to 1 congestion scenario, we add one server from each pod except pod 2, i.e., H3, H7, H15, then S9 is congested with a drop rate of 0.51\%.
There is no significant increase in drop rate and the number of drop location, as the congestion control mechanism of TCP Reno is taking effect.
For PABO, we can see that as the first aggregate switches each gathering traffic flows of  three connections, C1 and C4 bounce back packets to mitigate congestion. Since most of the congestion resulted from aggregating three connections is mitigate by the bouncing back of C1, C4. There is only a small scale of bounce back in S11 and no bounce back in S12.
The highest proportion of bouncing back is in S9 as it is the last hop that gathers all the traffic flows to H9.

In the 9 to 1 congestion scenario,we further add a server from each pod except pod 2, i.e., H2, H8 and H14.  In this scenario, pod switch S9 is congested with a drop rate of 0.45\% without PABO. 
When PABO is involved, as aggregate switches of five traffic flows from H1, H2, H5, H13 and H14, C1 and S11 bounce back packets.
For traffic flows from H3, H7, H8 and H15, the aggregate switch C4 bounces back packets. 
As the only path to H9, S9 still account for the highest percentage of bouncing back.
In total, 43.29\% of all the packets are bounced back.

In the 12 to 1 congestion scenario, all of the hosts in the first two pod and three out of four hosts (i.e., H13, H14, H15) in the last pod are the servers, together with a special intra-pod traffic brought by H11.
Without PABO, both pod switch S9 and core switch C1 are congested with a total drop rate of 0.52\%, while 32.6\% of the drop event happen in C1 and 67.4\% happen in S9. With PABO taking effect, C1, C4, S11, S12 and S9 bounce back 42.32\% of the packets.

For all the many-to-one scenarios, Table~\ref{tab:PassBackDistanceDistribution} shows that as the number of servers increases, the percentage of bounced packet decreases although the congestion is getting more severe. This is reasonable because though the percentage of bounced packets is smaller,  each bounced packet is bounced back and forth more frequently around the congestion point.

\begin{figure}
\centering
\includegraphics[scale=0.4]{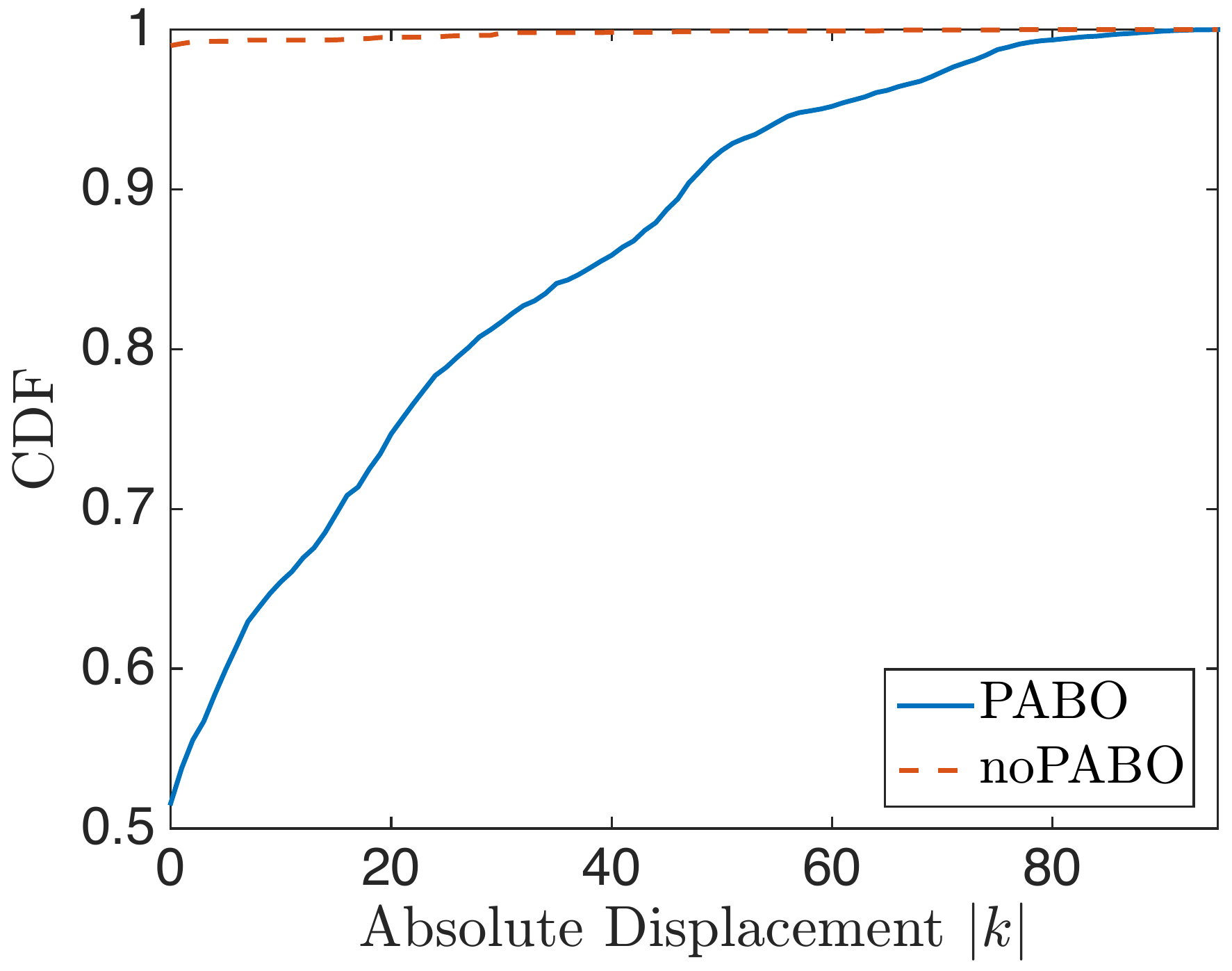}
\vspace{-0.3cm}
\caption{The CDF of absolute packet displacement in the 12 to 1 scenario. In cases without PABO, the percentage of out-of-order packets is much smaller,  but there are still packets with large displacement values,  which can trigger multiple times of retransmission timeout. }
\label{fig:rd_cdf}
\vspace{-0.5cm}
\end{figure}

Figure~\ref{fig:rd_cdf} shows the CDF of absolute packet displacement measured by RD in the 12 to 1 scenario.
In cases with PABO, because of the severe congestion, approximately 50\% of the packets arrive at their destination out of order. The max absolute value of displacement is 95.
In cases without PABO, packet loss is the main reason of packet out-of-order. Although the level of out-of-order is much less severe, the max absolute value of displacement is up to 76. This means there are still packets with large displacement values, which can trigger multiple times of retransmission timeout, resulting in a substantial delay stretch.

\begin{figure}
\centering
\includegraphics[scale=0.4]{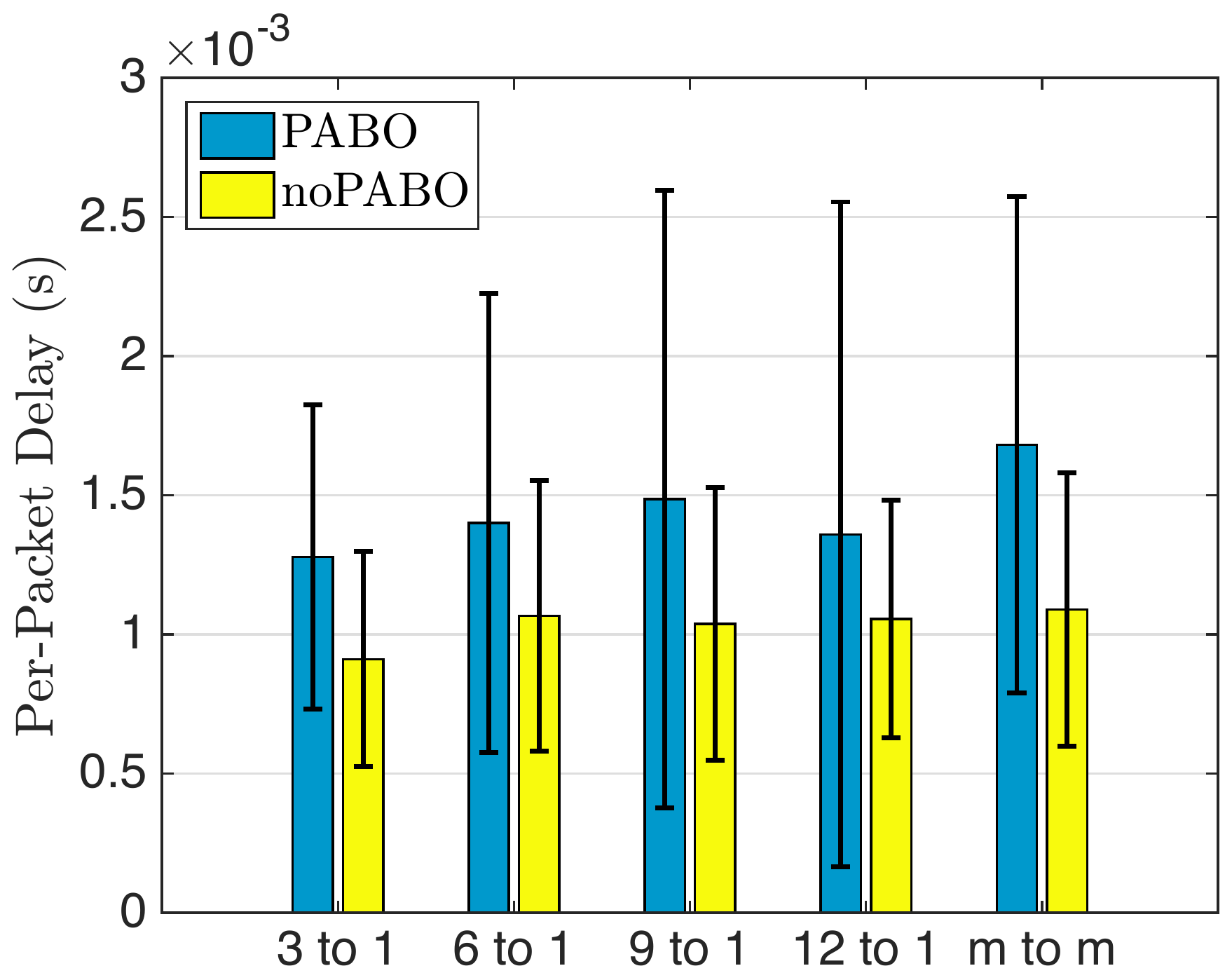}
\vspace{-0.3cm}
\caption{Average per-packet delay under different congestion scenario. The results with PABO is slightly higher than the results without PABO as PABO bounces back some of the packets.}
\label{fig:per_packet}
\vspace{-0.5cm}
\end{figure}

We also focus on the time delay comparison between cases with PABO and cases without PABO.
Figure~\ref{fig:per_packet} illustrates the average time delay per packet, regarding the time spent of every received packet from its source to destination. 
It shows that the average per-packet delay of PABO is slightly higher than cases without PABO, which is because some of the packets are bounced back to avoid packet loss.
Moreover, with the growth of the server number in the many-to-one scenario, the standard deviation of per-packet delay increases as the congestion is becoming more severe.
This confirm our statement that smaller percentage of bounced packets in more severe congestion means that bounced packets are bounced back and forth more frequently.
For all the many-to-one despite the 12 to 1 scenario, the basic trend for per-packet delay is very related to the degree of congestion. The exception is because the path of the special intra-pod traffic existed only in the 12 to 1 scenario is much shorter than the others. 

\begin{figure}
\centering
\includegraphics[scale=0.4]{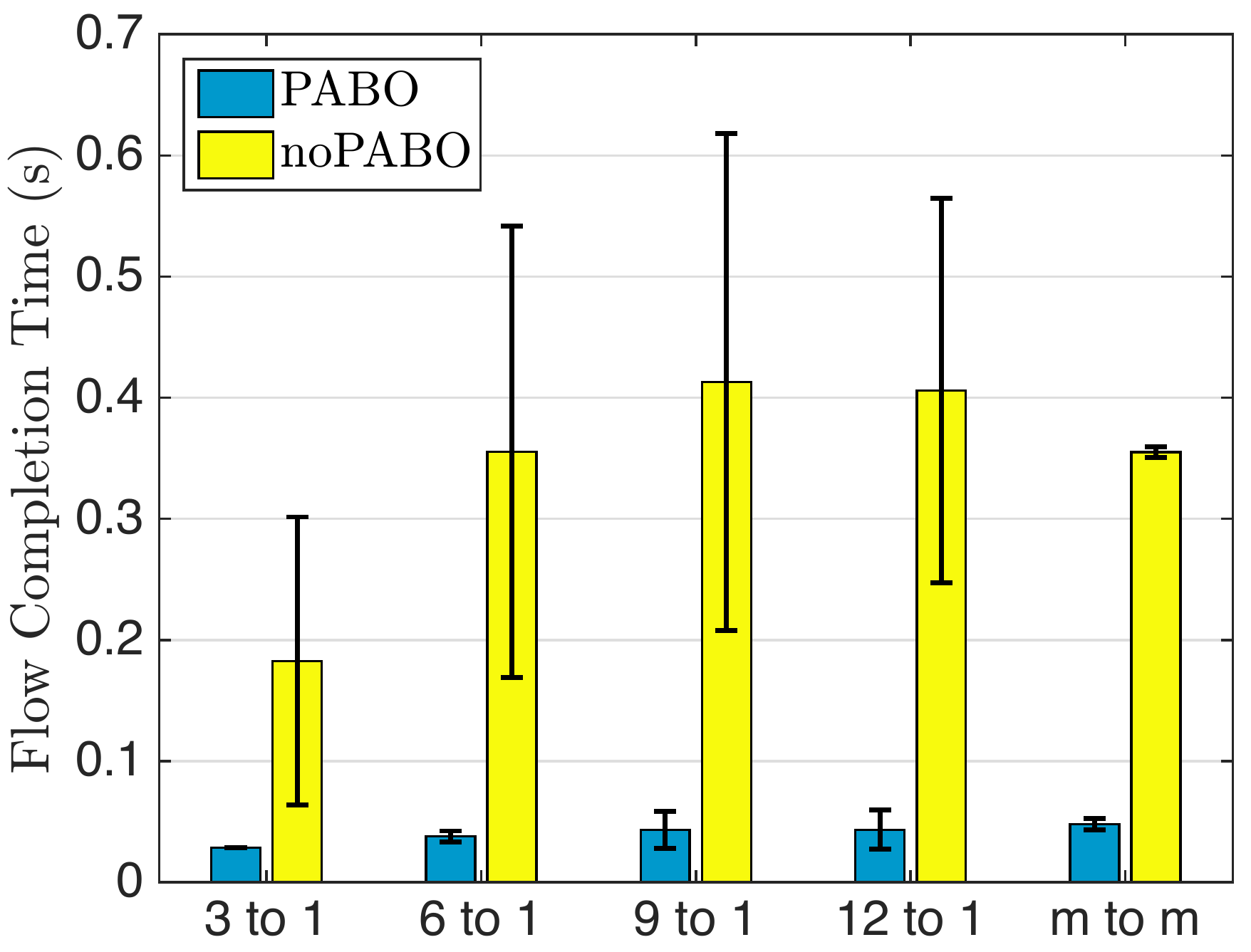}
\vspace{-0.3cm}
\caption{Average flow completion time in each congestion scenario. The results with PABO has obvious advantage over the results without PABO,  for packet losses leading to the orders of magnitude delay increases in retransmission-based approaches of TCP Reno.}
\label{fig:end_to_end}
\vspace{-0.5cm}
\end{figure}

Figure~\ref{fig:end_to_end} depicts the average flow completion time in each scenario,
which refers to the time taken from the client sends a request to its reception of the last packet of  the corresponding response.
From the figure, we can see that cases with PABO has obvious advantage over cases without PABO in all the scenarios, for retransmission-based approaches of TCP Reno bring orders of magnitude delay increases. Moreover, the advantage of PABO on flow completion time in the many-to-one scenario grows increasingly evident with the congestion becoming more severe (excluding the 12 to 1 scenario for the influence of  intra-pod traffic).

\subsection{Many-to-Many Scenario}
Second, we evaluate PABO under the many-to-many scenario, which consists of two 3 to 1 partition/aggregate traffic.
In this scenario, H1, H5 and H13 are the servers that respond to the request of client H9. In the meantime, H4, H8 and H16 respond to the request of client H10.
For each 3 to 1 traffic, the first aggregate switches are the core switch C1 and C2 respectively.
Then the two 3 to 1 traffic aggregate at S11 and go separately at the output ports of S9.
Without PABO, the drop rate is 0.54\%,  with all the packet losses occur in S11. 
By using PABO, 61.63\% of the packets are bounced back, with  18.35\% bounced at C1 ,  18.77\% at C2 and 62.88\% at S11.
In the many-to-many scenario,  the per-packet delay of PABO is still higher than cases without PABO, for over half of all the packets are bounced back. As for the results of flow completion time, PABO still has great advantage over the normal case.

\subsection{Impact of Parameter $\theta$}
\begin{figure}
\centering
\includegraphics[scale=0.4]{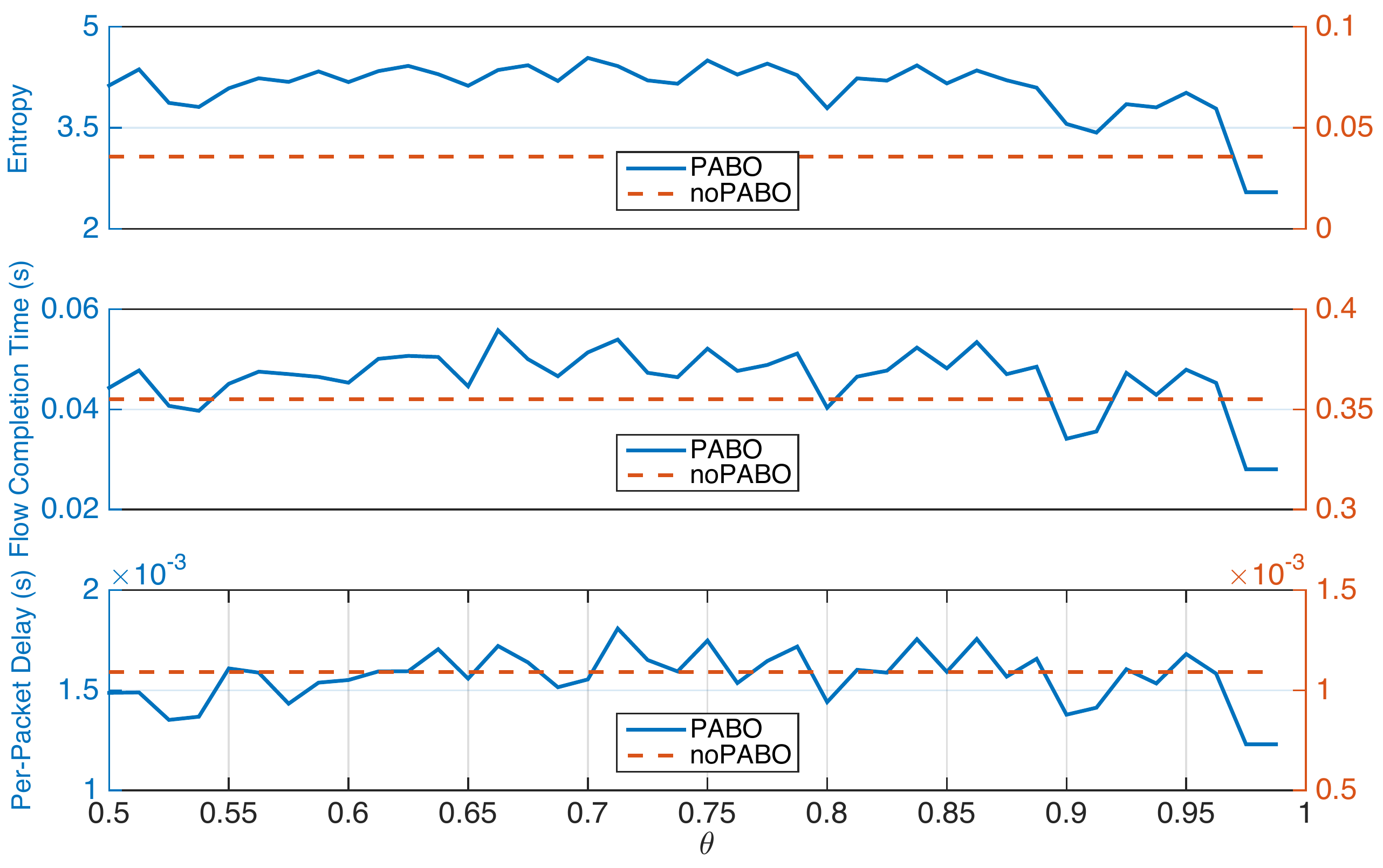}
\vspace{-0.3cm}
\caption{The relationship between $\theta$,  entropy, per-packet delay and flow completion time in the many-to-many scenario.}
\label{fig:theta_fattree}
\vspace{-0.5cm}
\end{figure}

We make evaluations to observe how the parameter $\theta$  in the bounce probability function of PABO affects on the experiment results under the many to many scenario.
We neglect the small values of $\theta$, because it's unnecessary to bounce back too early. Therefore, we only focus on the domain of $\theta \geq 0.5$.
The results are illustrated in Figure~\ref{fig:theta_fattree}.
It shows that with the increase of $\theta$, entropy remains stable then drops when $\theta$ is close to 1, which is similar to the previous result in Section~\ref{sec:out-of-order}.
As to the effect of  $\theta$ on the flow completion time, it shows no obvious regularity.
For no matter when PABO starts to bounce back packets, the bounced packets are absorbed inside the network to queue up for being finally handled by the destination end host. Therefore when using PABO,  the arrival time of the last packet is determined by the limit of the last hop switch connected to the destination, rather than the bouncing back threshold $\theta$.
For results of per-packet delay, there is also no obvious regularity. Smaller values of  $\theta$  avoid congestion at an earlier time, which result in larger percentage of bounced packets and relatively smaller bounced frequency for each packet. Similarly, larger $\theta$  values tend to avoid packet loss as well as maintaining high utilization of switches, thus the percentage of bounced packets is smaller and the bounced packets are bounced back and forth more frequently.

\section{Related Work}

We summarize some representative works on congestion control in data center networks and make a comparison with PABO in this section.


\bpara{Transport layer.} As the most frequently used transport layer protocol, TCP provides reliable end-to-end communication on unreliable infrastructures. Despite several variants of the traditional TCP protocol, the reactive fashion to congestion (i.e., timeout) and the slow-start nature in adjusting the sending window size cannot satisfy the growing requirements for small predictable latency and large sustained throughput in data center environments\cite{Alizadeh-DCTCP-2010}. 
ICTCP \cite{Wu-ICTCP-2010} aims at preventing incast congestion through adjusting the advertised windows sizes at the receiver side by estimating the available bandwidth and RTT. DCTCP \cite{Alizadeh-DCTCP-2010}, another TCP variant developed for data center environment, take advantage of the Explicit Congestion Notification (ECN) feature \cite{RFC3168-ECN} on switches to predict the extent of the congestion and provide smooth adjustments on the sending window size accordingly. These end-to-end solutions do not assume any reliability in the network and thus are orthogonal to PABO.

\bpara{Network layer.} AQM (Adaptive Queue Management) is an intelligent probabilistic packet dropping mechanism designed for switch buffer to avoid global synchronization among flows as can be frequently seen in traditional drop-tail queue settings (e.g., RED \cite{Floyd-RED-1993} and PI \cite{Hollot-PI-2001}). ECN (Explicit Congestion Notification) \cite{RFC3168-ECN} allows end-to-end congestion notification and ECN-enabled switches can set a marker (i.e., CE) in the IP header of the packet to signal impending congestion. This information will be echoed back to the sender with the ACK for this packet. While both can provide packet drop prevention at some degree, there is no guarantee on that no packet will be dropped. Fastlane \cite{Zats-Fastlane-2015} is an agile congestion notification mechanism, which aims at informing the sender as quickly as possible to throttle the transmission rate by sending explicit, high-priority drop notifications to the sender immediately at the congestion point.


\bpara{Link layer.} This research line aims at providing hop-by-hop reliability inside the network through a backpressure-alike feedback loop. Among them, PAUSE Frame \cite{Feuser_PFC_1999} is one of the flow control mechanisms for Ethernet, basing on the idea of sending PAUSE frame to halt the transmission of the sender for a specified period of time. However, PAUSE frame is per-link based and cannot differentiate among flows, which leads to performance collapse of all the flows on the link. PFC (Priority Flow Control)  \cite{Feuser_PFC_1999} further extends to provide individual flow control for several pre-defined service class. While it brings about some mitigation on inter-flow interference, the number of service class is still not enough in many circumstances. Moreover, the parameters of both PAUSE frame and PFC are very difficult to tune to ensure full reliability, making them unpractical \cite{Zats-DeTail-2012}.

DIBS \cite{Zarifis-DIBS-2014} solves local and transient congestions by detouring packets via a random port at the same switch when congestion occurs on a link. 
While adopting a similar idea of sharing switch buffers to mitigate transient congestion, PABO has three major additional merits: $i$) Packets are detoured by bouncing to upstream switches, which can minimize inter-flow interference on other paths, and is more manageable in maintaining packet order: bounced packets of a flow that queue up in the bounce queue are transmitted following First In First Out (FIFO) order along the same path.
$ii$) The bounced packets can also serve as a congestion notification for upstream nodes (switches or end-hosts).
$iii$) For considerations on reducing per-packet delay, the bounced packets are differentiated from normal packets and are limited in the number of times to be bounced.

\section{Conclusion}
In this paper, we proposed a reliable data transmission protocol -- PABO, for the link layer. When facing buffer saturation, PABO bounces the excess packets to upstream switches to avoid packet loss, which can mitigate transient congestion in network at a per-flow granularity. We complete a  proof-of-concept implementation, and investigate into the impact of PABO on the level of packet out-of-order, then we provide useful insights for configuring PABO. Extensive simulations have proved the effectiveness of PABO, showing that PABO has obvious superiority of flow completion time over the traditional protocol stack by guaranteeing zero packet loss in all cases. 

\ifCLASSOPTIONcaptionsoff
  \newpage
\fi



%
%



\bibliographystyle{IEEEtran}
\bibliography{IEEEabrv,ref}

%




\end{document}